\documentclass[12pt,a4papers,fleqn]{article}
\pdfoutput=1
\usepackage[english]{babel}

\usepackage{amssymb}
\usepackage{amsfonts}
\usepackage{amsmath}
\usepackage[round]{natbib}
\renewcommand{\cite}{\citet}
\usepackage[nohead]{geometry}
\usepackage[singlespacing]{setspace}
\usepackage[bottom]{footmisc}
\usepackage{indentfirst}
\usepackage{endnotes}
\usepackage[aboveskip=5pt]{caption}
\usepackage[skip=-20pt]{subcaption}
\usepackage{float}
\usepackage{graphicx}%
\usepackage{rotating}
\usepackage{verbatim}
\usepackage{setspace}
\usepackage{latexsym}
\usepackage{epstopdf}
\usepackage{float}
\usepackage{color}
\usepackage{booktabs}
\usepackage{longtable}
\usepackage{multirow}
\usepackage{bbm}

\usepackage[linesnumbered,ruled,vlined]{algorithm2e}
\SetKwInput{KwInput}{Step 1}
\SetKwInput{KwOutput}{Step 2}

\SetKwFunction{Forecasting}{Forecasting}
\usepackage{algpseudocode}

\usepackage{amsthm}
\usepackage{makeidx}
\usepackage{fancyhdr}
\usepackage{type1cm}
\usepackage{rotating}

\usepackage[pdfpagemode=UseOutlines ,plainpages=false
,hypertexnames=false ,pdfpagelabels ,hyperindex=true,colorlinks=true]{hyperref}
	\makeatletter
	\Hy@breaklinkstrue
	\makeatother
\usepackage{color}
\definecolor{darkred}{rgb}{0.6,0.0,0.1}
\definecolor{darkgreen}{rgb}{0,0.5,0}
\definecolor{darkblue}{rgb}{0,0,0.5}
\hypersetup{colorlinks ,linkcolor=darkblue ,filecolor=darkgreen
,urlcolor=darkblue ,citecolor=black}

\usepackage{url}

\RequirePackage{paralist}
\definecolor{dgreen}{rgb}{0,0.5,0}
\definecolor{dblue}{rgb}{0,0,0.9}
\definecolor{dred}{rgb}{0.6,0.0,0.1}
\definecolor{dgold}{rgb}{0.5,0.3,0.0}
\definecolor{dvio}{rgb}{0.6,0.3,0.5}
\definecolor{gray}{rgb}{0.5,0.5,0.5}
\definecolor{dbraun}{rgb}{.5,0.2,0}

\newcommand{\wtl}{\widetilde}
\newcommand{\wh}{\widehat}
\newcommand{\1}{\mathbbm{1}}
\newcommand{\E}{\mathbf{E}}
\newcommand{\ubeta}{\underline{\beta}}
\DeclareMathOperator*{\argmin}{argmin}

\newtheorem{cor}{Corollary}[section]

\usepackage{setspace}
\newtheoremstyle{mysc}% name
  {3pt}%      Space above
  {3pt}%      Space below
  {\it}%         Body font
  {}%         Indent amount (empty = no indent, \parindent = para indent)
  {\color{darkblue}\sc}% Thm head font
  {.}%        Punctuation after thm head
  {.5em}%     Space after thm head: " " = normal interword space;
  {}%         Thm head spec (can be left empty, meaning `normal')

\newtheoremstyle{myas}% name
  {3pt}%      Space above
  {5pt}%      Space below
  {\it}%         Body font
  {}%         Indent amount (empty = no indent, \parindent = para indent)
  {\color{darkblue}\sc}% Thm head font
  {.}%        Punctuation after thm head
  {.5em}%     Space after thm head: " " = normal interword space;
  {}%         Thm head spec (can be left empty, meaning `normal')

\newtheoremstyle{myex}% name
  {10pt}%      Space above
  {10pt}%      Space below
  {\it}%         Body font
  {}%         Indent amount (empty = no indent, \parindent = para indent)
  {\color{darkblue}\sc}% Thm head font
  {.}%        Punctuation after thm head
  {.5em}%     Space after thm head: " " = normal interword space;
  {}%         Thm head spec (can be left empty, meaning `normal')

\theoremstyle{mysc}\newtheorem{prop}{Proposition}[section]
\theoremstyle{mysc}
\theoremstyle{mysc}\newtheorem{tr}[prop]{Theorem}
\theoremstyle{mysc}

\theoremstyle{myas}\newtheorem{ass}{Assumption}

\theoremstyle{myex}
\theoremstyle{myex}

\numberwithin{equation}{section}

\newcounter{nc}
\makeatletter%
\def\@fnsymbol#1{\ensuremath{\ifcase#1\or 1 \or * \or  2\or 3\or 4\or * \or \star \or  , \or
g\or h\or i\else\@ctrerr\fi}}%
\makeatother%

\usepackage{xr}
\title{\vspace{-3cm} When are Google data useful to nowcast GDP? \\An approach via preselection and shrinkage
}

\author{\begin{minipage}{.29\textwidth}\center
{\sc Laurent Ferrara}\thanks{Skema Business School - University Cote d'Azur, and CAMA, Australian National University, e-mail:
\url{laurent.ferrara@skema.edu}}\\[.5ex]\small SKEMA Business School\\\null
\end{minipage}
\and \begin{minipage}{.28\textwidth}\center {\sc Anna
    Simoni}%\thanks{Corresponding  author.}
    \thanks{CREST, CNRS, ENSAE, \'{E}cole polytechnique - 5, avenue Henry Le Chatelier, 91120 Palaiseau, France,
e-mail: \url{anna.simoni@ensae.fr}}\\[.5ex]\small CREST, CNRS \\\null
\end{minipage}}

\begin{document}
\date{}
\maketitle

\begin{abstract}
{\footnotesize
Alternative data sets are widely used for macroeconomic nowcasting together with machine learning--based tools. The latter are often applied without a complete picture of their theoretical nowcasting properties. Against this background, this paper proposes a theoretically grounded nowcasting methodology that allows researchers to incorporate alternative Google Search Data (GSD) among the predictors and that combines targeted preselection, Ridge regularization, and Generalized Cross Validation. Breaking with most existing literature, which focuses on asymptotic in-sample theoretical properties, we establish the theoretical out-of-sample properties of our methodology and support them by Monte-Carlo simulations. We apply our methodology to GSD to nowcast GDP growth rate of several countries during various economic periods. Our empirical findings support the idea that GSD tend to increase nowcasting accuracy, even after controlling for official variables, but that the gain differs between periods of recessions and of macroeconomic stability.
}
\end{abstract}
{\footnotesize
\begin{tabbing}
\end{tabbing}}

% ====================================================================
\section{Introduction}\label{s:Introduction}
\indent Currently, practitioners widely use large sets of alternative data -- such as web--scraped data, Google data (\textit{i.e.} Trends, Correlate, or Search), scanner data, or satellite data --  for short-term macroeconomic forecasting and nowcasting purposes \citep[see \textit{e.g.}][]{Ng2017}. The main research questions related to alternative data sets are: (A) when such data improve nowcasting accuracy, and (B) whether they are useful even after controlling for official variables generally used by forecasters (\textit{e.g.} opinion surveys, production). Answering these questions requires the use of appropriate machine learning/econometrics methods. Recent macroeconomic empirical literature has seen an explosion of various methods to account for the specificity of alternative data, but for most of them we do not know their out-of-sample (OOS) theoretical properties, which matter the most for forecasting purposes. This paper puts forward a new methodology to deal with Google Search Data (GSD) for nowcasting purposes and establishes OOS large-sample properties for the proposed method. The challenging feature of GSD as a whole is their high dimension. GSD differ from Google Trends mainly in that GSD are volume variations of Google queries with respect to the first value while Google Trends provides the ratio between the search shares for a particular keyword/category over a given sub-period of time and the maximum search share for the same keyword/category over a larger chosen period. Macroeconomic nowcast based on GSD have been proposed in \textit{e.g.} \cite{GoetzKnetsch19} while Google Trends have been used in, for example, \cite{ChoiVarian09,ChoiVarian12}, \cite{ScottVarian15}, \cite{VosenSchmidt11}, \cite{DamuriMarcucci12}, \cite{NIESERTetal2020}, \cite{BorupSchutte2022}, and references therein. However, unlike our study, all these papers do not include theoretical contributions.\\
\indent Our proposed nowcasting method, which we call \textit{Ridge after Model Selection}, is a two-step approach: (i) first, GSD variables are preselected, conditional on the official variables, by targeting the macroeconomic aggregate to be nowcast, and (ii) second, a Ridge regularization is applied to those preselected GSD and official variables. The Ridge tuning parameter is chosen by Generalized Cross Validation (GCV). The main theoretical contributions of this paper are threefold. First, we prove that our targeted preselection retains all the variables in the true model with probability approaching one (sure screening property). Second, we establish an upper bound for both the in-sample and OOS prediction error associated with the Ridge after model selection estimator. Third, we evaluate optimality of GCV to choose the regularization parameter $\alpha$ of the Ridge regularization for OOS prediction. To the best of our knowledge, previous literature has established in-sample optimality of the GCV in the setting of Ridge regularization but not OOS optimality, see \textit{e.g.} \cite{Li1986} and \cite{CarrascoRossi2016}.\\
\indent \cite{Giannoneetal08} popularized the concept of macroeconomic nowcasting, which differs from standard forecasting approaches in the sense that it aims at evaluating current macroeconomic conditions on a high-frequency basis. The idea is to provide policy makers with a real-time evaluation of the state of the economy ahead of the release of official Quarterly National Accounts, which come out with a delay. For example, the New York Fed and the Atlanta Fed have recently developed new tools to evaluate US GDP quarterly growth on a high-frequency basis (see \url{https://www.newyorkfed.org/research/policy/nowcast} and \url{https://www.frbatlanta.org/cqer/research/gdpnow.aspx}). A large literature has developed that deals with nowcasting GDP growth for different countries, see \textit{e.g.} \cite{BoraganDieboldScotti2009}, \cite{DOZ2011}, \cite{AastveitTrovik12}, and \cite{FerraraMarsilli14}.\\
\indent To clarify the presentation of our nowcasting methodology, we present an example. Suppose the variable one wants to nowcast is the quarterly GDP growth rate. The predictors are made of two subsets: a set of official variables and a set of the weekly GSD variables. Our nowcasting model is based on linear regression models that incorporate predictors sampled over different frequencies (\textit{e.g.}, monthly and weekly) and released with various reporting lags so that the relevant information set for calculating the nowcast evolves within the quarter. To explicitly account for the information set available at a specific time within the quarter, a forecaster will consider a different set of predictors for each week, the same frequency of the higher-frequency variable considered in our example.\\
\indent To estimate the models, we use our aforementioned \textit{Ridge after Model Selection} procedure. The first step, preselection, is based on the t-statistics associated with each GSD variable in a regression that includes the official predictors as well, see \textit{e.g.} \cite{BaiNg2008} and \cite{BarutFanVerhasselt2016}. In the second step, a Ridge regularization is applied to the linear regression model incorporating the official variables and the GSD variables that have been preselected. The regularization parameter $\alpha$ is set equal to the minimum of the GCV criterion. Prior literature has proposed forecasting approaches based on Ridge regression to deal with dense models with a large number of predictors, \textit{e.g.} \cite{DEMOL2008} and \cite{CarrascoRossi2016}. We go beyond this literature by considering models in which the dimension can be ultra-high and that can be either sparse or dense.\\
\indent For our two-step procedure, we derive the three theoretical properties mentioned previously: the sure screening property of our preselection procedure, upper bounds for the in-sample and OOS prediction errors associated with the Ridge after model selection estimator, and optimality of the GCV for OOS prediction. The upper bounds are functions of the number of predictors $N$, the number of time-series observations $T$, and the Ridge regularization parameter $\alpha$. With regard to the GCV, we know from the previous literature that GCV has optimality properties for in-sample prediction, see \textit{e.g.} \cite{Li1986}, \cite{ANDREWS1991} and \cite{CarrascoRossi2016}. We complete this result by showing that the minimizer of the GCV is as good as the minimizer of the conditional mean squared prediction error for OOS prediction. The latter is the objective of nowcasting for central bankers.\\
\indent We study finite sample properties of our procedure using a Monte Carlo exercise. Our study analyzes how the dimension of the problem, $N$ and $T$, the degree of sparsity $s$ in the model, and the correlation among the predictors affect the performance of our method compared with other widely used methods in macroeconomic nowcast, like Lasso, Ridge without preselection, and Principal Component Analysis estimators. We show that when the true data generating process is sparse with a large number of active predictors, our \textit{Ridge after Model Selection} procedure outperforms all the considered competitors for OOS prediction.\\
\indent Finally, we conduct an empirical study to answer our two research questions (A) and (B) stated above for GSD with respect to GDP growth nowcast for three countries/areas: the euro area, the United States, and Germany. Common GDP nowcasting tools integrate standard official macroeconomic information stemming from, for instance, national statistical institutes, central banks, and international organizations. Two types of official data are typically considered: (i) hard data (production, sales, employment $\ldots$) and (ii) opinion surveys (households or companies are asked about their view on current and future economic conditions). Sometimes, financial markets information, generally available on high frequency basis, is also integrated into the information set. In our study, we include official data (i) and (ii) together with the alternative GSD into our information set. In addition, we consider financial market information for robustness check. \\
\indent We analyze three time periods: a period of cyclical stability ($2014q1-2016q1$), a period that exhibits a sharp downturn in GDP growth rate ($2017q1-2018q4$), and a period of recession (the Great Recession period from $2008q1$ to $2009q2$). Overall, empirical results show that GSD are useful when nowcasting GDP growth. At the beginning of the quarter, when no official information is available about the current state of the economy, we show that using only Google data leads to reasonable mean squared forecasting errors (MSFEs), sometimes only slightly higher than those obtained at the end of the quarter when the information set is complete. As soon as we integrate official macroeconomic information, starting from the fifth week of the quarter, MSFEs decrease, reflecting the importance of this type of data in nowcasting. Overall, combining macroeconomic variables and GSD variables in the same model appears to be fruitful.\\
\indent A striking result of our empirical analysis is that, on the one hand, the preselection step is crucial in the first two periods considered, as it generates better outcomes than nowcasting procedures that have no preselection. This result confirms previous findings from the nowcasting literature, see \textit{e.g.} \cite{BaiNg2008} and \cite{BoivinNg06} for dynamic factor models. On the other hand, recession periods present specific patterns as a model that only contains GSD, without any preselection step, tends to be preferred in terms of nowcasting accuracy. This result is quite robust over the three countries/areas that we consider in the study.\\
\indent The rest of the paper is organized as follows. In Section \ref{S:The_model}, we describe the nowcasting model and our \textit{Ridge after Model Selection} procedure, for which theoretical results are provided in this section and in Section \ref{S:theoretical:properties}. The Monte Carlo exercise is in Section \ref{ss:Monte:Carlo}, and the empirical application is in Section \ref{S:Empirical:Results}. Section \ref{S:Conclusions} concludes with a summary. Additional material and proofs are in the Supplementary Appendix.

% ====================================================================
\section{Methodology}\label{S:The_model}
\subsection{The nowcasting equation}\label{ss:nowcasting:approach}
To nowcast any series of interest $Y_t$, we focus on linear bridge equation models, which allow us to construct $Y_t$ nowcasts by using predictors available at different frequencies. To fix ideas, suppose the frequency of $Y_t$ is quarterly. We include three types of predictors: {\it soft } variables, such as opinion surveys, {\it hard } variables, such as industrial production or sales, and variables stemming from GSD. GSD are available on a weekly basis, and the other predictors are available on a monthly base. Let $t$ denote a given quarter of interest identified by its last month, for example the first quarter of $2005$ is dated by $t = \text{March} 2005$. The following is a bridge equation model to nowcast $Y_t$ for a specific quarter $t$, for $t=1,\ldots,T$:
  \begin{eqnarray}\label{eq_model}
    Y_t & = & \beta_0 + \beta_s'x_{t,s} + \beta_h'x_{t,h} + \beta_g'x_{t,g}  + \varepsilon_t, \qquad \E[\varepsilon_t| x_{t,s}, x_{t,h}, x_{t,g}] = 0,
  \end{eqnarray}
where $x_{t,s}$ is the $N_s$-vector containing {\it soft} variables, $x_{t,h}$ is the $N_h$-vector containing {\it hard} variables, $x_{t,g}$ is the $N_g$-vector of variables coming from GSD, and $\varepsilon_t$ is an unobservable shock. In general, $x_{t,s}$, $x_{t,h}$, and $x_{t,g}$ are sampled over different frequencies and released with various reporting lags, so the relevant information set for calculating the nowcasts evolves within the quarter. In practice, one must consider a model for each new available information set, which arises at the frequency of the highest frequency variable in the model, see Section \ref{SS:Google:data} for more details on this point. Model \eqref{eq_model} is estimated by the \emph{Ridge after model selection} procedure, which is summarized in Algorithm 1 and explained in details in Sections \ref{SS:pre_treatment} and \ref{SS:Ridge}.

\begin{algorithm}[!ht]\label{algorithm:1}
\footnotesize{
\KwData{Training sample $\{Y_t,x_{t,s},x_{t,h},x_{t,g}; t=1,\ldots,T\}$.}
\KwInput{}
\begin{itemize}
  \item[(1)] %For each $j = 1,\ldots, N_g$, regress $Y_t$ on a constant, $x_{t,s}$, $x_{t,h}$ and $x_{t,g,j}$, and compute the corresponding t-statistics $t_j$ associated with $x_{t,g,j}$;
  \For{$j = 1,\ldots, N_g$}{
    \textbf{regress} $Y_t$ on a constant, $x_{t,s}$, $x_{t,h}$, and $x_{t,g,j}$;\\
    \textbf{compute} the t-statistics $t_j$ associated with $x_{t,g,j}$;
  }
  \item[(2)] select the Google variables that have the largest absolute value $|t_j|$ compared with a given threshold $\lambda>0$: set $\wh{M}_g := \wh{M}_g(\lambda) := \left\{1\leq j\leq N_g: |t_j|>\lambda\right\}$.
\end{itemize}

\KwOutput{}
\begin{itemize}
  \item[(1)] Construct $\wh M =\wh M(\lambda) := \{1,\ldots,N_1\}\cup \{N_1 + j; j\in \wh M_g(\lambda)\}$, $x_{t,g,\wh M_g} := \{x_{t,g,j}; j\in \wh M_g\}$, and $X_{t,\wh M} := (1,x_{t,s}^{'},x_{t,h}^{'},x_{t,g,\wh M_g}^{'})'$;
  \item[(2)] for every $\alpha$ in a grid $\mathcal{A}$:
    \begin{itemize}
      \item[(a)] compute
        \begin{displaymath}
          \widehat\ubeta_{\wh M}(\alpha) = \left(\frac{1}{T}\sum_{t=1}^T X_{t,\wh M}X_{t,\wh M}^{'} + \alpha I_{\wh M}\right)^{-1}\frac{1}{T}\sum_{t=1}^T X_{t,\wh M }^{'}Y_t,
        \end{displaymath}
        \noindent where $I_{\wh M}$ is the $|\wh M|$-dimensional identity matrix;
      \item[(b)] compute
        \begin{displaymath}
          GCV(\alpha) := \frac{\sum_{t = 1}^T (Y_t - X_{t,\wh M}^{'}\widehat\ubeta_{\wh M}(\alpha))^2}{T\left(1 - tr(X_{t,\wh M}\left(T^{-1}\sum_{t=1}^T X_{t,\wh M}X_{t,\wh M}^{'} + \alpha I_{\wh M}\right)^{-1}X_{t,\wh M}^{'}/T)/T\right)^2},
        \end{displaymath}
        \noindent where $tr(\cdot)$ denotes the trace operator;
    \end{itemize}
  \item[(3)] select $\wh\alpha = \arg\min_{\alpha\in\mathcal{A}} GCV(\alpha)$.
\end{itemize}

%\Forecasting
\SetKwProg{Fn}{Forecasting}{:}{\KwRet}
\Fn{}{\KwData{$X_{\tau,\wh M}$ for $\tau > T$\;}
\KwRet $X_{\tau,\wh M}'\wh\ubeta_{\wh M}(\wh\alpha)$.}
\caption{Ridge after model selection}
}
\end{algorithm}
\normalsize

\subsection{Step 1: preselection of Google Search data}\label{SS:pre_treatment}
The recent literature on nowcasting and forecasting with large data sets concludes that using the largest available data set is not necessarily the optimal approach when nowcasting a specific macroeconomic variable such as GDP, at least in terms of nowcasting accuracy \citep[see \textit{e.g.}][]{BoivinNg06, Barhoumietal10}. The problem arises because we have too many variables and using all the variables would only add noise in the estimation process. As shown in \cite{BaiNg2008}, an empirical way to circumvent this issue is to more accurately target the choice of predictors.\\
\indent As we explain in Section \ref{S:Empirical:Results}, the GSD have a very high dimension $N_g$ compared with $T$, and in our empirical analysis we find that using all the variables in the GSD is not always a good strategy because we pay the price of dealing with ultra-high dimensionality without increasing the nowcasting accuracy as measured by the MSFE. In fact, it might be that the series $Y_t$ to be nowcast is highly predictable by a subset of the GSD and that this subset is specific to $Y_t$. For this reason, before estimating model \eqref{eq:model:weeks} we preselect GSD by retaining the most relevant variables for $Y_t$ nowcasting, capturing much of the variability in it. We refer to this approach as a targeted preselection.\\
\indent Preselection is based on the procedure proposed in \cite{BaiNg2008} and \cite{BarutFanVerhasselt2016} which, in our framework, works as follows. We begin with the standard linear regression equation \eqref{eq_model}. Then, by denoting with $x_{t,g,j}$ the $j$-th GSD variable, we apply the following approach:
\begin{itemize}
  \item[(1)] for each $j = 1,\ldots, N_g$, regress $Y_t$ on a constant, $x_{t,s}$, $x_{t,h}$, and $x_{t,g,j}$, and compute the corresponding t-statistics $t_j$ associated with $x_{t,g,j}$;
  \item[(2)] select the Google variables that have the largest absolute value $|t_j|$ compared with a given threshold $\lambda>0$: $\wh{M}_g := \wh{M}_g(\lambda) := \left\{1\leq j\leq N_g: |t_j|>\lambda\right\}$.
\end{itemize}
\noindent The basic idea of this approach is that a GSD variable is retained depending on its contribution for predicting $Y_t$ after controlling for a set of official variables. Associated with each $\lambda$ there is a selected submodel $\wh{M}_g := \wh{M}_g(\lambda)$. In practice, we take $\lambda$ as the $(1-\tau)$-quantile of a $\mathcal{N}(0,1)$ distribution with $\tau \in\{20\%, 10\%, 5\%, 2.5\%, 1\%,0.5\%\}$. The parameter $\tau$ must be interpreted as the percentage of false positives that can be tolerated. In contrast to \cite{BarutFanVerhasselt2016}, we do not assume that the conditional variance of $Y_t$ is known, and we estimate it to construct the t-statistics. This partially modifies the proof of the sure screening property in Theorem \ref{tr:sure:screening:property} below. This property is not established in \cite{BaiNg2008}.\\
\indent Let $N_1 := 1 + N_s + N_h$, $N := N_1 + N_g$, $X_t := (1,x_{t,s}', x_{t,h}', x_{t,g}')'$, $X_{t,O} := (1,x_{t,s}', x_{t,h}')'$, $X_{t,O,j} := (X_{t,O}', x_{t,g,j})'$ for every $j\in\{1,\ldots,N_g\}$, and $\beta := (\beta_0,\beta_s',\beta_h',\beta_g')'$. We introduce the following assumption.
\begin{ass}\label{Ass:1}
  Assume that : (i) $Y_t = \beta_*'X_t + \varepsilon_t$, $t = 1,\ldots,T$, with $\beta_*$ the true value of $\beta$, $\E[\varepsilon|X_t]=0$, and $\E[\varepsilon\varepsilon'|X_t] = \sigma^2 I$, where $\varepsilon = (\varepsilon_1,\ldots, \varepsilon_T)'$; (ii) $\beta_{*j} \neq 0$, $\forall j\leq N_1$ and for $1 \leq s_g^* \leq N_g$, $\beta_{*g} = (\beta_{*g,1}, \ldots,\beta_{*g,s_g^*},\mathbf{0}')'$, where $\mathbf{0}$ is a $(N_g - s_g^*)$-vector of zeros and $\beta_{*g,j}\neq 0$ for all $j=1,\ldots, s_g^*$; (iii) $\varepsilon_t|X_t$, $t\geq 1$ are independent zero-mean sub-Gaussian random variables.
\end{ass}
Assumption \ref{Ass:1} \textit{(i)} states that the true model is linear. Assumption \ref{Ass:1} \textit{(ii)} states that the subvector of the true $\beta_{*g}$ corresponding to the Google variables is $s_g^*$-sparse and that the true sparse model is $M^* := \{1,\ldots,N_1\}\cup \{N_1 + j; j\in M_g^*\}$, where $M_g^* := \lbrace 1 \leq j \leq N_g :\, \beta_{*g,j} \neq 0 \rbrace$ is the subset of the true sparse model containing only the indices of the active Google variables with size $s_g^* = | M_g^* |$. Assumption \ref{Ass:1} \textit{(iii)} assumes sub-Gaussianity of the errors conditional on the covariates $X_t$. This assumption is more general than assuming Gaussianity of $\varepsilon_t|X_t$ and allows for distributions whose tails are dominated by the tails of a Gaussian distribution.\\
\indent For every $j\in\{1,\ldots,N_g\}$, define $\wtl\beta_{O,j} := \arg\min_{\beta_{O,j}^1,\beta_{g,j}}\E(Y_t - X_{t,O}'\beta_{O,j}^1 - x_{t,g,j}\beta_{g,j})^2$, which is the pseudo-true value of $\beta_{O,j} := (\beta_{O,j}^{1'},\beta_{g,j})'\in\mathbb{R}^{N_1+1}$ in the $j$-th misspecified model and define $\sigma_{O,j}^2 := \E[(Y_t - X_{t,O,j}'\wtl\beta_{O,j})^2]$. Misspecification arises because, in general, $\wtl{\beta}_{O,j}$ differs from the corresponding coefficients in $\beta_{*}$. Finally, $\mathcal{B} := \{\beta_{O,j}, j= 1,\ldots, N_g;|\beta_{O,j,1}|\leq B, \ldots ,|\beta_{O,j,N_1}|\leq B, |\beta_{O,j,N_1+1}|\leq B\}$ for a large positive constant $B$ is the set over which the Least Squares estimates in step (1) are searched. The next assumption allows us to control for the estimated $\sigma_{O,j}^2$ in the construction of the t-statistics.
\begin{ass}\label{Ass:homoskedasticity}
  Assume that: \textit{(i)} there exists a constant $c>0$ such that $\E[\|x_{t,g}'\beta_{*g}\|_2^2] \leq c$; \textit{(ii)} $\{(x_{t,s}', x_{t,h}', x_{t,g}')'\}_{t \geq 1}$ is a zero mean strictly stationary sequence with values in $\mathbb{R}^{N-1}$; \textit{(iii)} for every $j \in\{1, \ldots, N_g\}$, $\E[\wtl\varepsilon_{t,j}^4| X_{t,O,j}]$ is bounded, where $\wtl\varepsilon_{t,j} := (y_t - X_{t,O,j}'\wtl\beta_{O,j})$; \textit{(iv)} there exist two constants $0 < \underline{C}_x^2 < \overline{C}_x^2 < \infty$ such that
  $$\underline{C}_x^2 \leq \min_{1 \leq j \leq N_g} \lambda_{\min}\left(\E[X_{t,O,j}X_{t,O,j}']\right) \leq \max_{1 \leq j\leq N_g} \lambda_{\max}\left(\E[X_{t,O,j}X_{t,O,j}']\right) \leq \overline{C}_x^2,$$ where $\lambda_{\min}(\cdot)$ and $\lambda_{\max}(\cdot)$ denote the minimum and maximum eigenvalues of a matrix; \textit{(v)} there exist two constants $0< \underline{\sigma}_{O}^2 < \overline{\sigma}_{O}^2 < \infty$ such that $\underline{\sigma}_{O}^2 \leq \min_{1 \leq j \leq N_g}\sigma_{O,j}^2 \leq \max_{1 \leq j \leq N_g}\sigma_{O,j}^2 \leq \overline{\sigma}_{O}^2$.
\end{ass}
\noindent The first three parts of the next assumption are the same as Conditions 1 and 2 in \cite{BarutFanVerhasselt2016} but made specific to our framework; we refer to that paper for comments about it. Parts \textit{(iv)} and \textit{(v)} of the next assumption allow us to control the deviation of an empirical process without assuming independency. In Sections C.5.1--C.5.2 of the Supplementary Appendix we verify parts \textit{(iv)} and \textit{(v)} of Assumption \ref{Ass:0} for the i.i.d. case. For a function $h$, define $\mathbb{G}_T [g(u_t)] := \frac{1}{\sqrt{T}}\sum_{t=1}^T [g(u_t) - \E(g(u_t))]$.
\begin{ass}\label{Ass:0}
  Assume that: \textit{(i)} for $j\in M_g^*$, there exist two positive constants $c_1, c_2 >0$ and $0<\kappa < 1/2$ such that $|cov(Y_t,x_{t,g,j}|X_{t,O})| \geq c_1 T^{-\kappa}$,
  and uniformly in $j\in\{1,\ldots, N_g\}$: $\E[x_{t,g,j}^2] \leq c_2$; \textit{(ii)} there exists a sufficiently large constant $\kappa_T$ such that for $\epsilon_T := 16 \kappa_T K_T (1 + \ell)\sqrt{(N_1 + 1)/(T\underline{C}_x^4)}$ with $\underline{C}_x^2$ given in Assumption \ref{Ass:homoskedasticity} (iv), $\ell$ a positive constant, and $K_T$ an arbitrarily large constant, it holds for all $j\in\{1, \ldots, N_g\}$:
  $$\sup_{\beta\in\mathcal{B}; \|\beta - \wtl\beta_{O,j}\| \leq \epsilon_T}\left|\E\left(\left[\left(Y_t - X_{t,O,j}'\beta\right)^2 - \left(Y_t - X_{t,O,j}'\wtl \beta_{O,j}\right)^2\right]\1_{\Omega_{T,O,j}^c}(X_{t,O,j},Y_t)\right)\right| \leq o(N_1/T),$$
  \noindent where $\Omega_{T,O,j}^c = \{|Y_t| > m_0 K_T^{\rho}/s_0\} \,\bigcup_{l=1}^{N_1 + 1}\{|X_{t,O,j,l}| > K_T\}$; \textit{(iii)} there exist positive constants $m_0, m_1, s_0, s_1$, and $\rho$ such that for sufficiently large $\tau$, $P\left(|X_{t,j}|> \tau \right) \leq m_1\exp\{-m_0 \tau^{\rho}\}$ for all $j\in\{2, \ldots, N\}$, and $\E[\exp\{2\beta_*'X_t s_0\} + \exp\{-2\beta_*'X_t s_0\}] \leq s_1$; \textit{(iv)} for every $\ell>0$,
  \begin{multline*}
    P\Big(\sup_{\beta\in\mathcal{B}; \|\beta - \wtl\beta_{O,j}\| \leq \epsilon_T}\Big|(\beta + \wtl{\beta}_{O,j})'\frac{1}{\sqrt{T}}\mathbb{G}_T\left[X_{t,O,j}X_{t,O,j}'\1_{\Omega_{T,O,j}}\right](\beta - \wtl{\beta}_{O,j})\\
     - 2\frac{1}{\sqrt{T}}\mathbb{G}_T\left[Y_t X_{t,O,j}'\1_{\Omega_{T,O,j}}\right](\beta - \wtl{\beta}_{O,j})\Big|\geq 4\epsilon_T \kappa_T K_T \sqrt{\frac{N_1 + 1}{T}} \,\ell\Big)\leq e^{ - 2\ell^2},
  \end{multline*}
  \noindent where $\epsilon_T$ is defined as previously and $\1_{\Omega_{T,O,j}} := \1_{\Omega_{T,O,j}}(X_{t,O,j}, Y_t)$ with $\Omega_{T,O,j} := \{\|X_{t,O,j}\|_{\infty} \leq K_T, |Y_t| \leq m_0 K_T^{\rho}/s_0\}$, $\|\cdot\|_{\infty}$ the maximum norm, and $m_0, \rho, K_T$ as introduced previously; \textit{(v)} for every positive constants $\ell, b_1 $, and $K_T$ the arbitrarily large constant introduced previously and $\1_{T,O,j} := \1\{\|X_{t,O,j}\|_{\infty} \leq K_T\}$:
  \footnotesize{
  \begin{multline*}
    P\left(\left.\left\|\frac{1}{T}\sum_{t=1}^T\left(X_{t,O,j}X_{t,O,j}' - \E[X_{t,O,j}X_{t,O,j}'\1_{T,O,j}]\right) \right\|_{op} > b_1 T^{- \ell}\right|\Omega_{T,O,j}\right)\\
    \leq 2(N_1+1) \exp\left\{- \frac{T^{1 - 2\ell} b_1^2}{4(N_1 + 1)^2 K_T^4\max\{b_1,1\}}\right\}.
  \end{multline*}
  }
\end{ass}
The next theorem establishes the sure screening property of our selection procedure.
\begin{tr}\label{tr:sure:screening:property}
  Suppose that Assumptions \ref{Ass:1}--\ref{Ass:0} hold. Let $\kappa_T := K_TB(N_1 + 1) + m_0K_T^{\rho}/s_0$, with $K_T$ and $\rho$ given in Assumption \ref{Ass:0} (ii)--(iii). Assume that $T^{1 - 2\kappa}/(\kappa_T K_T)^2 \rightarrow \infty$ and that $T^{-\kappa/2}K_T^{\rho/2} =\mathcal{O}(1)$ with $\kappa < 1/2$ given in Assumption \ref{Ass:0} \textit{(i)}. Then, by taking $\lambda = c_0 T^{1/2 -\kappa}$ for some constant $c_0>0$, it holds that
  \begin{multline*}
    P\left( M_g^* \subset \wh{M}_g(\lambda)\right) \geq  1 - 8s_g^*(N_1 + 1)\exp\left\{-\frac{\min\{c_2,b_1^2/4,1/4\}}{\kappa_T^2 K_T^2} T^{1 -2\kappa}\right\}\\
    -6s_g^*Tm_2e^{-m_0 K_T^{\rho}} - 2s_g^* T \exp\left\{- \frac{K_T^{\rho}}{4C K_1^2}\right\} - 2 s_g^* \exp\left\{ - c_1 T^{1 - 2\kappa} \min\left\{\frac{c_{\epsilon}^2}{4K_1^2},\frac{c_{\epsilon}}{2K_1}\right\}\right\},%\\
  \end{multline*}
  \noindent where $b_1, C,c_{\epsilon}$ are positive constants, $c_2 := \underline{C}_x^2 c_1^2/(256 N_1)$, $m_2 := (N_1m_1 + \sqrt{s_1} \sqrt{\E[\exp\{4C_m s_0^2\|\varepsilon\|_{\psi_2}^2\}]})$ for some positive constants $c_1,\, C_m$ and with $\|\varepsilon\|_{\psi_2} := \max_t\sup_{p\geq1} p^{-1/2} \left(\E\left[\left.|\varepsilon_t|^p\,\right|X_t\right]\right)^{1/p}$, $K_1 := \max_{j\in M_g^*}\max_{t}\|\wtl\varepsilon_{t,j}^2\|_{\psi_1}$, with $\|\cdot\|_{\psi_1}$ denoting the sub-exponential norm.
\end{tr}
We present a proof of this theorem in Appendix \ref{Appendix:proof:Sure:Screening}. The result of the theorem is similar to \cite[Theorem 3]{BarutFanVerhasselt2016}. The difference, between our result (and proof) and theirs, is the presence of additional terms in the lower bound of the probability. These terms are present in our result due to the variance estimation in our approach (to construct the t-statistic $t_j$ and $\wh M_g$). Instead, \cite{BarutFanVerhasselt2016} assume the variance to be known. If $1/(4CK_1^2)\geq m_0$, $\log(s_g^*) = o(\min\{T^{1 - 2\kappa}/(\kappa_T K_T)^2, K_T^{\rho}, T^{1 - 2\kappa}/(\max\{K_1^2,K_1\})\})$, $\log(N_1)\frac{(\kappa_T K_T)^2}{T^{1 - 2\kappa}} < \min\{c_2,b_1^2/4,1/4\}$, and $\log(T m_2) < m_0K_T^{\rho}$, then Theorem \ref{tr:sure:screening:property} establishes that the selected submodel includes the true model $M_g^*$ with probability approaching one. Moreover, if $\underline{c}<\min\left\{\frac{c_{\epsilon}^2}{K_1^2},\frac{c_{\epsilon}}{K_1}\right\}< \overline{c}$, for two positive constants $\underline{c},\overline{c}$, then the last two terms in the rate are negligible with respect to the other ones, and if we take the optimal order $K_T\asymp T^{(1 - 2\kappa)/A}$, where $A := \max\{4 + \rho, 2 + 3\rho\}$, then $P\left( M_g^* \subset \wh{M}_g(\lambda)\right) \gtrsim  1 - s_g^* T m_2 \exp\{- C T^{(1 - 2\kappa)\rho/A}\}$. In this case, it follows from Lemma B.1 in the Supplementary Appendix that with our methodology we can deal with an $N_g$ such that $\log(N_g) = o(T^{(1 - 2\kappa)\rho/A})$. Similarly, we can deal with an $s_g^*$ and an $m_2$ such that: $\log(s_g^*) = o(T^{(1 - 2\kappa)\rho/A})$ and $\log(m_2) = o(T^{(1 - 2\kappa)\rho/A})$, which means that $N_1$ and $\|\varepsilon\|_{\psi_2}$ are allowed to increase at a certain rate. If $X_{t,j}$ are sub-Gaussian, then Assumption \ref{Ass:homoskedasticity} \textit{(iv)} is satisfied with $\rho = 2$, which gives $\log(N_g) = o(T^{(1 - 2\kappa)/4})$.
% ====================================================================

\subsection{Step 2: Ridge regression}\label{SS:Ridge}
\indent Because GSD have a very high dimension, with the number of variables being much larger than the number of observations, even after the preselection in Step 1 the number of selected Google variables may still be large compared with the time dimension $T$. To deal with this large number of preselected covariates, in Step 2 we use Ridge regularization. Let $\wh M =\wh M(\lambda) := \{1,\ldots,N_1\}\cup \{N_1 + j; j\in \wh M_g(\lambda)\}$ and denote $X_{t,\wh M} := (1,x_{t,s}^{'},x_{t,h}^{'},x_{t,g,\wh M_g}^{'})'$, where $x_{t,g,\wh M_g} := \{x_{t,g,j}; j\in \wh M_g\}$ is the vector containing only the preselected Google variables. We estimate the parameter $\beta$ in equation \eqref{eq_model} by minimizing a penalized residuals sum of squares in which the penalty is given by the squared Euclidean norm $\|\cdot\|_2$, and we define the \emph{Ridge after Model Selection} estimator as: $\widehat\beta := \widehat\beta(\alpha)$, where
\begin{equation}\label{Ridge:after:preselection}
  \widehat\beta(\alpha) := \argmin_{\beta\in\mathbb{R}^N; \beta_{g,j}=0, j\in \wh M_{g}^c}\left\{\frac{1}{T}\sum_{t = 1}^T\left(Y_t - \beta_0 - \beta_s'x_{t,s} - \beta_h'x_{t,h} - \beta_g'x_{t,g}\right)^2 + \alpha \|\beta\|_2^2\right\},
\end{equation}
\noindent and $\alpha > 0$ is a regularization parameter that tunes the amount of shrinkage. Without loss of generality, we can assume that the selected elements of $x_{t,g}$ corresponding to the indices in $\wh M_g$ are the first elements of the vector. Let $\mathbf{0}$ be the $(N_g - |\wh M|)$-dimensional column vector of zeros. Then, we can write $\widehat\beta$ as $\widehat\beta := (\widehat\ubeta_{\wh M}^{'}, \mathbf{0}')'$, where
\begin{displaymath}
  \widehat\ubeta_{\wh M} = \widehat\ubeta_{\wh M}(\alpha) = \left(\frac{1}{T}\sum_{t=1}^T X_{t,\wh M}X_{t,\wh M}^{'} + \alpha I_{\wh M}\right)^{-1}\frac{1}{T}\sum_{t=1}^T X_{t,\wh M }^{'}Y_t,
\end{displaymath}
\noindent and $I_{\wh M}$ is the $|\wh M|$-dimensional identity matrix with $|\wh M| = N_1 + |\wh M_g|$.\\
\indent Empirical choice of the parameter $\alpha$ is based on the generalized cross-validation (GCV) technique \citep[see][]{Li1986,Li1987}, the idea of which is to choose a value for $\alpha$ for which the MSFE is as small as possible. \cite{CarrascoRossi2016} recently used this technique in an in-sample forecasting setting. To complement their study, we show in Section \ref{ss_optimality_GCV} that GCV performs well for out-of-sample prediction as well. With GCV, the researcher selects the value of $\alpha$ that minimizes the following quantity:
\begin{displaymath}
  GCV(\alpha) := \frac{\sum_{t = 1}^T (Y_t - X_{t,\wh M}^{'}\widehat\ubeta_{\wh M}(\alpha))^2}{T\left(1 - tr(X_{t,\wh M}\left(T^{-1}\sum_{t=1}^T X_{t,\wh M}X_{t,\wh M}^{'} + \alpha I_{\wh M}\right)^{-1}X_{t,\wh M}^{'}/T)/T\right)^2},
\end{displaymath}
\noindent where $T$ denotes the number of quarters in the training sample and $tr(\cdot)$ denotes the trace operator. We denote by $\wh\alpha$ the value of $\alpha$ that minimizes $GCV(\alpha)$. In the next section, we establish the theoretical properties of $\widehat\beta$ and $\wh\alpha$.

% ====================================================================
%\input{Asymptotic}
\section{Theoretical Properties}\label{S:theoretical:properties}
\subsection{In-sample and Out-of-sample Prediction Error}\label{ss:prediction:error}
In this section, we study the convergence to zero of the in-sample and OOS prediction error associated with the \emph{Ridge after Model Selection estimator}. Asymptotic properties for the OOS prediction error associated with the Ridge estimator without model selection have been analysed in \cite{DEMOL2008} and \cite{CarrascoRossi2016} for dense models while asymptotic properties for the in-sample prediction error are well known in the inverse problems literature; for example see \cite{CarrascoFlorensRenault2007} and for a Bayesian interpretation of the Ridge estimator see \cite{FlorensSimoni2012,FS14} .\\
\indent To the best of our knowledge, extant literature has not established the theoretical properties of the Ridge estimator coupled with a targeted selection. Here, we fill this gap and establish an upper bound for both the in-sample and OOS prediction error for sparse models. This upper bound gives the rate of convergence as $N,T\rightarrow \infty$. It also gives the rate for dense models -- that is, when $s_g^* = N_g$.\\
\indent Let $X := (X_1,\ldots,X_T)'$ be a $(T\times N)$ matrix. Let $M^{*c}$ denote the complementary set of $M^*:= \lbrace 1 \leq j \leq N : \beta_{*j} \neq 0 \rbrace$ in $\{1,\ldots,N\}$ with $s^* := |M^*|$. For a vector $\beta\in\mathbb{R}^N$ and an index set $M\subset\{1,\ldots,N\}$, denote $\beta_M := (\beta_{M,j})_{j=1}^N$ with $\beta_{M,j} := \beta_j\1\{j\in M\}$, and for a $(T\times N)$ matrix $X$, denote by $X_M$ the $(T\times |M|)$ matrix made of the columns of $X$ corresponding to the indices in $M$, and denote by $X_{t,M}$ the transpose of the $t$-th row of $X_M$. Thus, $\beta_M$ has zero outside the set $M$. %We denote by $P_{X}$ the conditional probability given the covariates $X$.
For a vector $\delta\in\mathbb{R}^N$ and given covariates $X_t$, $t=1,\ldots, T$, define the squared prediction norm of $\delta$ as $\|\delta\|_{2,T}^2 := \delta'X'X \delta/T$, the $\ell_0$-norm of $\delta$ as $\|\delta\|_0 := \sum_{j=1}^N \1\{\delta_j\neq 0\}$ and the Euclidean norm as $\|\delta\|_2 := \sqrt{\delta'\delta}$.\\
\indent Next, we introduce an assumption known in the literature as a restricted sparse eigenvalue condition on the empirical Gram matrix $X'X/T$ \citep[see \textit{e.g.},][]{belloni2013}, and it is an extension of the restricted isometry condition \citep[\textit{e.g.},][]{candesTao2007}. The quantity $m$ in the assumption restricts the number of nonzero components outside the set $M^*$ of the vectors $\delta\in\mathbb{R}^N$ considered. The larger $m$ is, the more restrictive the first part of the assumption is.
\begin{ass}\label{Ass:3}
  For a given $m<T$, for a $\delta\in\mathbb{R}^N$, with probability $1 - o(1)$,
  $\underline\varphi(m)^2 := \min_{\|\delta_{M^{*c}}\|_0 \leq m,\,\delta \neq 0}\frac{\|\delta\|_{2,T}^2}{\|\delta\|_2^2} > 0$ and $ \overline\varphi(m) := \max_{\|\delta_{M^{*c}}\|_0\leq m, \,\delta\neq 0} \frac{\|\delta\|_{2,T}^2}{\|\delta\|_2^2} > 0.$
\end{ass}
\noindent \cite[Theorem B.2]{ChernozhukovHardleHuangWang2021} provide primitive conditions on the population covariance matrix ensuring that this assumption holds for covariates $X_t$ with temporal dependence. We define the condition number associated with the empirical Gram matrix $(X_{\wh M}'X_{\wh M})/T$:
$\mu(\wh m) = \frac{\sqrt{\overline\varphi(\wh m)}}{\underline\varphi(\wh m)}$, where $\wh m:= |\wh M \setminus M^*| \, \1\{\wh M \supseteq M^*\}$ is the number of incorrect covariates selected. Similarly, define $\wh k := |M^* \setminus \wh M| \, \1\{M^*\nsubseteq \wh M\}$. We start by establishing an upper bound on the in-sample prediction error. For a proof see the Supplementary Appendix C.1.
\begin{tr}[In-sample prediction error]\label{thr:3}
  Suppose that Assumptions \ref{Ass:1} \textit{(i)}--\textit{(ii)} and \ref{Ass:3} are satisfied and that $\varepsilon_t|X_t$ is Gaussian. Let $\wh M$ be the model selected in the first step. Let $\wh \beta$ be the Ridge estimator defined in equation \eqref{Ridge:after:preselection}. Then, for every $\epsilon > 0$, there is a constant $K_{\epsilon}$ independent of $T$ such that with probability at least $1 - \epsilon$,
  \footnotesize{
  \begin{multline*}
    \|\wh\beta - \beta_*\|_{2,T} \leq \left(K_{\epsilon}\sqrt{\frac{(\wh m + s^*)[\log(N) + \log(e^2\mu(\wh m))]}{T}} + 2\alpha \|\beta_{*}\|_2 \frac{1}{\underline\varphi(\wh m)}\right)\1\{M^*\subseteq \wh M \}+\\
    \left(\frac{K_{\epsilon}\sigma}{\sqrt{T}}\sqrt{(\wh k + \wh m)\left(\log(s^* + \wh m) + \log(e^2\mu(\wh k + \wh m))\right)} + \frac{2\alpha}{\underline\varphi(\wh m)} \|\beta_{*}\|_2 + \|\beta_{*,M^*\setminus\wh M}\|_{2,T}\right)\1\{M^*\nsubseteq \wh M \}.
  \end{multline*}
  }
\end{tr}
\indent As noted in the discussion following Theorem \ref{tr:sure:screening:property}, $P(M^* \subset \wh M) \rightarrow 1$ under some conditions (and so is the probability of $\{M^*\subseteq \wh M \}$). We remark that on the event $\{M^*\nsubseteq \wh M\}$, instead, we get a bias term given by $\|\beta_{*,M^*\setminus\wh M}\|_{2,T}$, which is intuitive because the second-step Ridge estimator is always biased for the components in $M^*\setminus \wh M$.\\
\indent The next corollary establishes an upper bound for the Euclidean norm of $(\wh\beta - \beta_*)$.
\begin{cor}[Coefficient estimation]\label{cor:1}
  Suppose that Assumptions \ref{Ass:1} \textit{(i)}--\textit{(ii)} and \ref{Ass:3} are satisfied and that $\varepsilon_t|X_t$ is Gaussian. Let $\wh M$ be the model selected in the first step. Let $\wh \beta$ be the Ridge estimator defined in equation \eqref{Ridge:after:preselection}. Then, for every $\epsilon > 0$, there is a constant $K_{\epsilon}$ independent of $T$ such that with probability at least $1 - \epsilon$,
  \footnotesize{
  \begin{multline*}
    \|\wh\beta - \beta_*\|_{2} \leq \left(K_{\epsilon}\sqrt{\frac{(\wh m + s^*)[\log(N) + \log(e^2\mu(\wh m))]}{T\underline\varphi(\wh m)^2}} + 2\alpha \|\beta_{*}\|_2 \frac{1}{\underline\varphi(\wh m)^2}\right)\1\{M^*\subseteq \wh M \} + \\
    \left(\frac{K_{\epsilon}\sigma}{\underline\varphi(\wh m)\sqrt{T}}\sqrt{(\wh k + \wh m)\left(\log(s^* + \wh m) + \log(e^2\mu(\wh k + \wh m))\right)} + \frac{2\alpha}{\underline\varphi(\wh m)^2} \|\beta_{*}\|_2 + \frac{\|\beta_{*,M^*\setminus\wh M}\|_{2,T}}{\underline\varphi(\wh m)}\right)\1\{M^*\nsubseteq \wh M \}.
  \end{multline*}
  }
\end{cor}
Compared with the upper bound for the in-sample prediction error, every term in the upper bound in Corollary \ref{cor:1} has an additional factor of $1/\underline\varphi(\wh m)$. As seen in Assumption \ref{Ass:3}, $\underline\varphi(\wh m)$ must be interpreted as the smallest restricted eigenvalue of the empirical Gram matrix and so it can be small when $N$ is large. Therefore, the upper bound in Corollary \ref{cor:1} can be larger than the upper bound in Theorem \ref{thr:3}.\\
\indent In the next theorem, we establish an upper bound for the OOS prediction error. Let $(Y_{\tau},X_{\tau}')'$, $\tau > T$, be a new copy of $(Y_t,X_t')'$ that satisfies Assumption \ref{Ass:1} \textit{(i)}-\textit{(ii)} and that is independent of $(Y,X)$ with $Y:=(Y_1,\ldots,Y_T)'$.
\begin{cor}[OOS prediction error]\label{cor:2}
  Let the assumptions and the notations of Theorems \ref{tr:sure:screening:property} and \ref{thr:3} %and of Lemma \ref{lem:B:9}
  hold. Let $\wh \beta$ be the Ridge estimator defined in equation \eqref{Ridge:after:preselection}. Take $K_T \asymp T^{(1 - 2\kappa)/A}$, where $A := \max\{4 + \rho, 2 + 3\rho\}$, and let $c<\min\left\{\frac{c_{\epsilon}^2}{K_1^2},\frac{c_{\epsilon}}{K_1}\right\}< C$ %and $0<c<K_2<C$
  for two positive constants $c,C$, $\log(N_g m_2) = o(T^{(1 - 2\kappa)\rho/A})$. Then there is a constant $K_{\epsilon}$ independent of $T$ such that with probability converging to $1$, conditional on $X_{\tau}$,
  \begin{multline*}
    X_{\tau}'(\wh \beta - \beta_*) \leq \|X_{\tau,\wh M}\|_2\left(K_{\epsilon}\sqrt{\frac{(\wh m + s^*)[\log(N) + \log(e^2\mu(\wh m))]}{T\underline\varphi(\wh m)^2}} + 2\alpha \|\beta_{*}\|_2 \frac{1}{\underline\varphi(\wh m)^2}\right).
  \end{multline*}
\end{cor}
The upper bound for the OOS prediction error is larger than the upper bound for the in-sample prediction error because $X_{\tau}$ has dimension $N$, which is large. Preselection allows us to reduce this dimension from $N$ to $(\wh m + s^*)$, and we do not need to assume that $\|X_{\tau}\|_2 = O_p(1)$ as, for example, in \cite{CarrascoRossi2016}.

\subsection{Out-of-sample evaluation of the selection of $\alpha$}\label{ss_optimality_GCV}
In this section, we evaluate the performance of $\wh\alpha$, the minimizer of the GCV as described in Section \ref{SS:Ridge}, for OOS prediction which is the objective of nowcasting for central bankers. Previous research \citep[]{Li1986,CarrascoRossi2016} has established optimality of GCV minimization for in-sample prediction in the setting of Ridge regularization, but these studies do not consider optimality for OOS prediction. Although \cite{Leeb2008} does so in a setting different from Ridge regularization to evaluate the performance of model selection our proofs depart entirely from their proofs.\\
\indent Consider a new copy $(Y_{T+1},X_{T+1}')'$ of $(Y_t,X_t')'$ that satisfies Assumption \ref{Ass:1} \textit{(i)}--\textit{(ii)} and that is independent of $(Y^{(T)},X^{(T)})$, where $Y^{(T)}:=(Y_1,\ldots, Y_T)'$ and $X^{(T)} := (X_1, \ldots, X_T)'$, that is, $Y_{T+1} = \sum_{j=1}^{s^*} X_{T+1,j}\beta_{*,j} + \varepsilon_{T+1} = \sum_{j=1}^{\wh m + s^*} X_{T+1,j}\beta_{*,j} + \varepsilon_{T+1}$ with $\beta_{*,j} = 0$ for every $j \in\{s^* + 1, \ldots, \wh m + s^*\}$, $\E[\varepsilon_{T+1}|X_{T+1}] = 0$, and $Var(\varepsilon_{T+1}|X_{T+1}) = \sigma^2$. When researchers aim to forecast, they want that the $\wh\alpha$ selected by GCV (based on the sample $(Y, X)$) be optimal for OOS prediction. The OOS performance of a selected value $\wh\alpha$ is evaluated by considering the conditional mean squared prediction error given by
$$\rho^2(\alpha;Y^{(T)},X^{(T)}) := \E[(Y_{T+1} - \wh\beta(\alpha)'X_{T+1})^2|Y^{(T)},X^{(T)},\mathcal{A}],$$
where $\wh\beta(\alpha)'X_{T+1} = \sum_{j=1}^{\wh m + s^*}X_{T+1,j}\wh\beta_j(\alpha)$, $\wh\beta(\alpha)$ is the \emph{Ridge after Model Selection} estimator defined in equation \eqref{Ridge:after:preselection}, and $\mathcal{A} := \{M^*\subseteq \wh M\}$. Theorem \ref{thr:4} provides the rate at which the GCV criterion converges to $\rho^2(\alpha;Y^{(T)},X^{(T)})$ uniformly over $\alpha$ in a given region as $T\rightarrow \infty$. To this end, we introduce the following notation: for a vector $\beta\in \mathbb{R}^N$ and an index set $M\subset \{1,\ldots, N\}$, denote $\ubeta_M := (\beta_{j})_{j\in M}\in\mathbb{R}^{|M|}$, $\wh\Sigma_{ M}:= X_{M}'X_{M}/T$, $\Sigma_{M} := \E[X_{t,M} X_{t,M}']$, and $\mathcal{B}(M) := \left\{X^{(T)};\left\|\wh\Sigma_{M} - \Sigma_{M}\right\|_{op} \leq C\sqrt{\frac{|M|}{T}}\right\}$ with $0<C<\infty$ as a universal constant. For a matrix $A$, $\|A\|_{op}$ denotes its operator norm.
\begin{ass}\label{Ass:4}
  Assume that: \textit{(i)} $\E[\varepsilon_t^4] < C_1$ for some constant $0<C_1<\infty$; \textit{(ii)} for every $t$, $\E[X_{t,\wh M} X_{t,\wh M}'|X^{(t-1)},\mathcal{A}] = \Sigma_{\wh M}$, and $\Sigma_{\wh M}$ is bounded almost surely; (iii) $P(M^*\nsubseteq \wh M) = o(r_{\mathcal{A},T})$, where $r_{\mathcal{A},T}$ is a non-stochastic sequence independent of $\alpha$ that converges to zero as $T\rightarrow \infty$; (iv) $P(\mathcal{B}(\wh M )^c) = o(r_{\mathcal{B},T})$, where $r_{\mathcal{B},T}$ is a non-stochastic sequence independent of $\alpha$ that converges to zero as $T\rightarrow \infty$;
  and (v) for any index set $M\subset \{1,\ldots, N\}$, there exist a $w(M)\in\mathbb{R}^{|M|}$ and $\gamma > 0$ such that $\ubeta_{*,M} = \Sigma_{M}^{\gamma/2} w(M)$ and $\|w(M)\|_2 < \infty$ (source condition).
\end{ass}

Assumption \ref{Ass:4} \textit{(iii)} is satisfied by our preselection method in Step 1 under Assumptions \ref{Ass:1}--\ref{Ass:0}. Theorem \ref{tr:sure:screening:property} gives an explicit expression for $r_{\mathcal{A},T}$ under those assumptions. Assumption \ref{Ass:4} \textit{(iv)} is satisfied, for example, if $X_{t,\wh M}$ are independent sub-Gaussian random vectors with sub-Gaussian norm bounded by a constant $K$. In this case, $C$ in the definition of $\mathcal{B}$ is equal to $4 K^2$ and $r_{\mathcal{B},T} = 2\E[\exp\{2\log(\wh m + s^*) - c \min\{\sqrt{\frac{\wh{m} + s^*}{T}},\frac{\wh{m} + s^*}{T}\}T\}]$ for a numerical constant $c>0$ by the Bernstein's inequality for sub-exponential random variables \citep[see, \textit{e.g.} ][Theorem 1.4.1]{vershynin2018four}.
\begin{tr}\label{thr:4}
  Let Assumptions \ref{Ass:1} \textit{(i)}--\textit{(ii)} and \ref{Ass:4} hold. Let $\varepsilon_t|X_t$, $t\geq 1$ be independent zero-mean random variables, $(\wh m + s^*)/T < 1$ for every $T\geq 1$ with probability $1$, and $\wtl\gamma := \min\{\gamma,2\}$. Then, for every $\alpha>0$,
  \begin{multline*}
    \left|\rho^2(\alpha;Y^{(T)},X^{(T)})  - GCV(\alpha)\right|\\
    = \mathcal{O}_p\left(\alpha^{\wtl\gamma} + \frac{1}{\alpha \sqrt{T}} + \frac{1}{\sqrt{T}} + \frac{(\wh m + s^*)}{T}\alpha^{(\wtl\gamma - 2)}\right) + \mathcal{O}_p\left(\max\{r_{\mathcal{A},T},r_{\mathcal{B},T}\}\right).
  \end{multline*}
  Moreover, for any constants $0<\underline{\alpha}<\infty$ and $0<\underline{u}<1/2$, and for a sequence $\overline{\alpha}_T\rightarrow 0$ as $T\rightarrow \infty$ such that: $T^{-1/2 + \underline{u}}/\overline{\alpha}_T \rightarrow 0$, it holds that
  \begin{displaymath}
    \sup_{\alpha\in[\underline{\alpha}T^{-1/2 + \underline{u}},\overline{\alpha}_T]}\left|\rho^2(\alpha;Y^{(T)},X^{(T)})  - GCV(\alpha)\right|
    = \mathcal{O}_p\left(r_T\right) + \mathcal{O}_p\left(\max\{r_{\mathcal{A},T},r_{\mathcal{B},T}\}\right),
  \end{displaymath}
  \noindent where $r_T := \overline{\alpha}_T^{\wtl\gamma} + T^{-\underline{u}} + \frac{(\wh m + s^*)}{T} T^{(2 - \wtl\gamma)(1 - 2\underline{u})/2}$.
\end{tr}
\noindent This theorem establishes the rate at which the conditional mean squared prediction error converges to the $GCV$ criterion. If $\frac{(\wh m + s^*)}{T^{2\underline{u}} T^{\wtl\gamma(1/2 - \underline{u})}} \rightarrow 0$, then the convergence of the two criteria is uniform over a shrinking interval. The next theorem establishes the OOS optimality of the GCV-minimizer $\wh\alpha$.

\begin{tr}\label{thr:5}
  Consider the minimizers of $GCV(\alpha)$ and $\rho^2(\alpha;Y^{(T)},X^{(T)})$: $$\wh\alpha = \arg\min_{\alpha\in[\underline{\alpha}T^{-1/2 + \underline{u}},\overline{\alpha}_T]} GCV(\alpha)$$
  \noindent and $\alpha^* = \arg\min_{\alpha\in[\underline{\alpha}T^{-1/2 + \underline{u}},\overline{\alpha}_T]}\rho^2(\alpha;Y^{(T)},X^{(T)})$, where $\underline{\alpha}$, $\underline{u}$, and $\overline{\alpha}_T$ are as defined in Theorem \ref{thr:4}. Then, in the setting of Theorem \ref{thr:4}, (i) $\wh\alpha$ is as good as $\alpha^*$ for OOS prediction in the sense that
  \begin{displaymath}
    |\rho^2(\wh\alpha;Y^{(T)},X^{(T)}) - \rho^2(\alpha^*;Y^{(T)},X^{(T)})| = \mathcal{O}_p\left(r_T\right) + \mathcal{O}_p\left(\max\{r_{\mathcal{A},T},r_{\mathcal{B},T}\}\right),
  \end{displaymath}
  \noindent where $r_T$ is defined in Theorem \ref{thr:4}, and (ii) the OOS predictive performance can be consistently estimated in the sense that
  \begin{displaymath}
    |GCV(\wh\alpha) - \rho^2(\wh\alpha;Y^{(T)},X^{(T)})| = \mathcal{O}_p\left(r_T\right) + \mathcal{O}_p\left(\max\{r_{\mathcal{A},T},r_{\mathcal{B},T}\}\right).
  \end{displaymath}
\end{tr}
% ====================================================================
\section{Monte Carlo exercise}\label{ss:Monte:Carlo}
This section presents the results of a simulation exercise. We are interested in understanding how the dimension of the problem, $N$ and $T$, the degree of sparsity $s^*$ in the model, and the correlation among the predictors affect the nowcasting performance of our estimation method in finite sample. To this end, we conduct two exercises. The first consists in comparing our \emph{Ridge after Model Selection} procedure with the most commonly used methods in the macroeconometric nowcasting/forecasting literature. For the description and results of this exercise, see Section D in the Supplementary Appendix. Here, we show the results of the second exercise, in which we examine the effect of varying $N,T,s^*$ on the in-sample and OOS prediction error to confirm the theoretical results in Section \ref{ss:prediction:error}.\\
\indent The data are simulated according to the following DGP: $t=1,\ldots, T$,
\begin{eqnarray}
  y_t & = & \gamma'z_t + \beta'x_t + v_t,  \qquad z_t = (z_{1,t},z_{2,t})'\sim \mathcal{N}_2\left(0,\left[\begin{array}{cc}
    1 & 0.3\\
    0.3 & 1
  \end{array}\right]\right), \nonumber\\
  \underset{(N\times 1)}{x_t} & = & \delta'z_t + u_t,
\end{eqnarray}
\noindent where $\gamma = (1,2)'$, $u_t\sim \mathcal{N}_N(0,\Psi)$, $\Psi$ is an $(N\times N)$-full rank covariance matrix, and $v_t\sim\mathcal{N}(0,1)$. In the DGP for $y_t$, we have two sets of covariates: $z_t$, which we are sure is in the model, and $x_t$, which must be preselected. The possible sparsity of the model only affects $x_t$. Specification of $\beta$ governs the sparsity of the model, $\delta$ determines the correlation between $x_t$ and $z_t$, and $\Psi$ affects the correlation among covariates in $x_t$ that are included in the model and those that are not. We consider a sparse structure: $\beta_j\sim\mathcal{N}(0,1)$ for $j\leq s^*$ and $\beta_j = 0$ for $j> s^*.$\\
\indent For the parameter vector $\delta$, we use the specifications: $\delta = 0.2\iota$ and $\delta = 0.8\iota$ with $\iota$ denoting a $(2\times N)$ matrix of ones. For the covariance matrix $\Psi$, we consider two cases: (I) \textit{uncorrelated:} $\Psi = I_N$ and (II) \textit{decreasing correlation}: $\Psi = \left(|0.5|^{j-k}\right)_{j,k}$. For $N,T,s$, we consider: $(N=150, T=100,s=105)$, $(N=200, T=150,s=105)$, $(N=200, T=150,s=110)$, and $(N=200, T=100,s=110)$. We adjust split between in-sample and OOS to keep the same fraction of observations in the two samples.\\
\indent We present the results in Tables \ref{Table:Results:NTs:paper:1}-\ref{Table:Results:NTs:paper:2} in Appendix \ref{s:Appendix:paper} as ratios with respect to the case $N=150, T=100,s=105$. That is, we show the in-sample mean squared error (MSE) of each triplet $(N,T,s)$ relative to the in-sample MSE for $N=150, T=100,s=105$, labeled ``MSER'', and the OOS MSE of each triplet $(N,T,s)$ relative to the OOS MSE for $N=150, T=100,s=105$, labeled ``MSFER''. The threshold parameter $\lambda$ is set equal to the $\{80\%, 90\%,95\%,97.5\%,99\%, 99.5\%\}$-quantiles of a $\mathcal{N}(0,1)$ distribution and is indicated in the first column of Tables \ref{Table:Results:NTs:paper:1} and \ref{Table:Results:NTs:paper:2}. This choice corresponds to a false discovery rate (FDR) equal to $20\%, 10\%, 5\%, 2.5\%, 1\%, 0.5\%$, respectively.\\
\indent We see that when $T$ increases from $100$ to $150$, both the MSE (in-sample error) and the MSFE (OOS error) decrease even if $N$ increases as well (first two columns in Tables \ref{Table:Results:NTs:paper:1}-\ref{Table:Results:NTs:paper:2}). When $s$ is also increasing (going from $s=105$ to $s=110$), we see a reduction in the MSE and MSFE, even if it is smaller (third and fourth columns in Tables \ref{Table:Results:NTs:paper:1}-\ref{Table:Results:NTs:paper:2}), and for the values of $\lambda$ corresponding to a FDR of $20\%$ (and also $10\%$ in the case $\delta = 0.2\iota$), only the MSFE decreases. The conclusions are similar for the two structures of $\Psi$. Instead, when $N$ and $s$ increase but $T$ remains fixed (last two columns in Tables \ref{Table:Results:NTs:paper:1}-\ref{Table:Results:NTs:paper:2}), the MSE is decreasing in most of the cases but the MSFE is increasing. This illustrates our extra term $\|X_{\tau,\wh M}\|_2$ in the upper bound for the OOS prediction error in Corollary \ref{cor:2}.

% ====================================================================
\section{Empirical Study}\label{S:Empirical:Results}
In this section, we present the empirical results obtained by applying our \emph{Ridge after Model selection} procedure to nowcast GDP growth rate with weekly GSD for three countries/areas: the euro area (EA, hereafter), the United States, and Germany. We present our OOS results for three various phases of the business cycle: a calm period (2014--2016), a period with a sudden downward shift in GDP growth (2017--2018) and a recession period with large negative growth rates (2008--2009). We consider 2014--2016 a calm period because it does not show any strong GDP movements, excepting a decline in oil prices starting from mid-2014. The period 2017--2018 is interesting in that although the global economy was recovering in 2017 at a faster pace than economists expected the trade war initiated by the Trump administration came by surprise, leading to a sharp slowdown in global GDP growth amidst rising uncertainties around global trade. Last, all the considered countries/areas experienced a large drop in the level of GDP during the Great Recession of 2008--2009, in the wake of the global financial crisis. During these three periods, the quality and sample size of GSD may have improved, for instance, due to the population's improved access to internet. While we implicitly assume that no structural change in the quality of GSD occurred over the sample considered, interpretation of our empirical findings should take this assumption into account.\\
\indent We compute the empirical results against a background of pseudo real-time analysis; that is, we account for the release dates of official variables, but we do not use vintages of data.

\subsection{Design of the empirical analysis}\label{SS:Google:data}
The objective of this empirical application is to nowcast on a high-frequency basis quarterly GDP growth rate (variable $Y_t$ in model \eqref{eq_model}) for three countries/areas. Official GDP data come from Eurostat for EA as a whole, from Destatis for Germany, and from the Bureau of Economic Analysis for the U.S.
The official monthly macroeconomic series that we use as regressors $x_{t,h}$ and $x_{t,s}$ are, respectively, the growth rate of the industrial production index, which is the measure of hard data most used by practitioners, denoted by $IP_t$, and a composite index of opinion surveys from various sectors as a proxy for soft variables,  denoted by $S_t$. As regards Germany and EA, we use for $S_t$ sentiment indexes computed by the European Commission, and we use the well known ISM index for the U.S. economy.

GSD are weekly data received and made available by the European Central Bank every Tuesday. GSD are data related to queries performed with Google search machines. Google assigns queries to particular categories using natural language processing methods. GSD are not the same as Google Trends data but rather are indexes of weekly volume changes of Google queries grouped by category and by country. Data are normalized at $1$ at the first week of January 2004, the first week of their availability. Then, the following values indicate the deviation from the first value. The GSD we use for our study cover weekly Google searches for the six main EA countries: Belgium, France, Germany, Italy, the Netherlands, and Spain, as well as for the United States, ranging from the January 1, 2004, to December 31, 2018. For each country, we have at our disposal a total of $N_g = 296$ variables (categories). When dealing with the EA as the whole, we account for information conveyed by the six main countries, meaning that we have access to a total of $N_g = 1,776$ GSD variables.

Treating weekly data is particularly challenging as the number of entire weeks present in every quarter is not always the same, and a careful analysis must be done when incorporating these data. Original data are not seasonally adjusted, thus we take the growth rate over 52 weeks to eliminate the seasonality within the data, a standard procedure when dealing with weekly data \citep[see][]{Lewisetal20}. To account for the variation over a quarter, we add a second differentiation over 13 weeks. Consequently, models are estimated on a recursive basis starting the last week of March 2005.

In addition to the challenge of frequency mismatch in the data, we must address the fact that data on official series and GSD are released with various reporting lags, which leads to an unbalanced information set at each point in time within the quarter. In the literature, this issue is referred to as a {\it ragged-edge database} \citep[see][]{Angelinietal11}.\\
\indent Because of the different frequencies of $x_{t,s}$, $x_{t,h}$, and $x_{t,g}$ -- monthly and weekly, respectively -- and of the various reporting lags, the relevant information set for calculating the nowcasts evolves within the quarter. We assume that a given quarter is made up of thirteen weeks. Thus, by letting $x_{t,j,(v)}$ designate the vector of series in $x_{t,j}$ released at week $v$ of period $t$ and $x_{t,j,(v),i}$ the $i$-th series in $x_{t,j,(v)}$, we denote by $x_{t,j,i}^{(w)}$ a summary of $\{x_{t,j,(v),i}, 1\leq v\leq w\}$. Then, we
%Thus, by denoting the $i$-th series in vector $x_{t,j}$ released at week less or equal than $w=1,\ldots,13$ of quarter $t$ as $x_{t,j,i}^{(w)}$, $j \in \{s,h,g\} $, we
define the relevant information set at week $w$ of a quarter $t$ as $\Omega_t^{(w)} := \{x_{t,j,i}^{(w)}, i=1,\ldots, N_j,\, j\in\{s,h,g\}\}.$ For simplicity, we keep in $\Omega_t^{(w)}$ only the observations relative to the current quarter $t$ and do not consider past observations. While the series in $x_{t,g}$ are in $\Omega_t^{(w)}$ for every $w=1,\ldots,13$, the other variables are in the relevant information set only for the weeks corresponding to (and after) their release, and so the data set is unbalanced.\\
\indent To explicitly account for the relevant information set at each week of the quarter, we replace model \eqref{eq_model} by a model for each week $w$ denoted by $M_{(w)}$ and defined as: $\forall t=1,\ldots,T$, $\forall w=1,\ldots,13$,
  \begin{eqnarray}\label{eq:model:weeks}
    M_{(w)}:\qquad
    \E[Y_t|\Omega_t^{(w)}] & = & \beta_{0,w} + \beta_{s,w}'x_{t,s}^{(w)} + \beta_{h,w}'x_{t,h}^{(w)} + \beta_{g,w}'x_{t,g}^{(w)},
  \end{eqnarray}
\noindent where $x_{t,g}^{(w)}:= (x_{t,g,i}^{(w)})_{i=1,\ldots,N_g}$, $\beta_{j,w,i} = 0$ if $x_{t,j,i}^{(w)}\notin \Omega_t^{(w)}$. For instance, as the first observation of EA-industrial production relative to the current quarter $t$ is released in week $9$, we set the corresponding $\beta_{h,w} = 0$ for every $w=1,\ldots, 8$. The bridge equation \eqref{eq:model:weeks} exploits weekly information to obtain more accurate nowcasts of $Y_t$. The idea of having $13$ models is that a researcher aiming to nowcast the current-quarter values of $Y_t$ will use the model corresponding to the current week of the quarter. For instance, to nowcast the current-quarter value of $Y_t$ at the end of week $2$, we would use model $M_{(2)}$.\\
\indent With regards to macroeconomic series, we mimic the exact release dates as published by statistical offices. Within the EA, for instance, the first survey of the quarter, referring to the first month, typically arrives in week 5. Then, the second survey of the quarter, related to the second month, is available in week 9. The $IP_t$ for the first month of the quarter is available about 45 days after the end of the reference month, which is generally in week 11. Finally, the last survey, related to the third month of the quarter, is available in week 13 (see Figure 1 in the Supplementary Appendix). For the U.S. economy, survey data are released at the same dates as for the EA, but industrial production data are released about four weeks earlier.

To construct the vector $x_{t,g}^{(w)}$ in equation \eqref{eq:model:weeks} containing GSD variables, we take for each Google variable the sample average of its observations over the period week $1$ to week $w$ of quarter $t$. That is, %by denoting with $x_{t,g,(v)}$ the $N_g$-vector of Google variables released at week $v$ of period $t$ (not averaged),
we construct $x_{t,g}^{(w)} = \sum_{v=1}^{w} x_{t,g,(v)}/w$. Take, for instance, $w =3$ (\textit{i.e.} model $3$ which is used at week $3$), then $x_{t,g}^{(3)}$ is equal to $(x_{t,g,(1)} + x_{t,g,(2)} + x_{t,g,(3)})/3$.\\
\indent For the survey $S_t$ in $x_{t,s}^{(w)}$ and the industrial production $IP_t$ in $x_{t,h}^{(w)}$, we impose the following structure, which mimics the data release explained previously in the case of the EA. The variable $x_{t,s}^{(w)}$ is not present in models $1$ to $4$ because the current $S_t$ is not available in the first four weeks of the quarter, so that $\beta_{s,1} = \beta_{s,2} = \beta_{s,3} = \beta_{s,4} = 0$. Then, for models $w\in\{5,\ldots,8\}$, $x_{t,s}^{(w)}$ is the value of $S_t$ for the first month of the quarter: $x_{t,s}^{(w)} = S_{t,1}$, where $S_{t,i}$ denotes the variable $S_t$ referring to the $i$-th month of quarter $t$. In models $w\in\{9,\ldots,12\}$, $x_{t,s}^{(w)}$ is equal to the average of the survey data available at the end of the first and second months of the quarter: $x_{t,s}^{(w)}= (S_{t,1} + S_{t,2})/2$. Lastly, in model $13$, $x_{t,s}^{(w)}$ is the average of the survey data over the quarter: $x_{t,s}^{(w)}= (S_{t,1} + S_{t,2} + S_{t,3})/3$.
Similarly, the variable $x_{t,h}^{(w)}$ is not present in models $w\in\{1,\ldots,10\}$ (so that $\beta_{h,1} = \ldots = \beta_{h,10} =0$), and in models $w\in\{11,\ldots,13\}$, $x_{t,h}^{(w)}$ is the value of $IP_{t}$ for the first month of quarter $t$. For Germany and the United States we use a similar structure that mimics the specific data release for these countries.

\indent
We split our data set in two non-overlapping subsamples: the training set and the OOS set. The latter starts at $2014q1$, $2017q1$, or $2008q1$, depending on the period we are considering, and the training sample always finishes two quarters before the beginning of the OOS period to take into account the delay due to the release of GDP figures (see Section E.5 in the Supplementary Appendix). We use a recursive scheme method, that is, we re-estimate the parameters at each new nowcasting quarter using all the past information available until the penultimate quarter before the nowcasting one.\\
\indent In the implementation of the preselection step of our procedure, we can take into account the problem of frequency mismatch among the different sets of variables and make each monthly/weekly series comparable to the quarterly $Y_t$ series in terms of frequency by replacing each predictor with either its value available at a given week during the quarter $t$ or with the average of its observations over the quarter $t$. In the second case, we replace $x_{t,s}$, $x_{t,h}$, and $x_{t,g}$ by $\sum_{m = 1}^{3}x_{t,s,(m)}/3$, $\sum_{m = 1}^{3}x_{t,h,(m)}/3$, and $x_{t,g}^{(13)} := \sum_{w = 1}^{13}x_{t,g,(w)}/13$, respectively, with $x_{t,s,(m)}$, $x_{t,h,(m)}$ and $x_{t,g,(v)}$ denoting the vectors of soft, hard, and Google variables released at month $m$ and week $v$ of quarter $t$, respectively. In this application, we carry out the preselection step only once by using the training sample available for the first nowcast origin and the average values of the observations on the covariates over each quarter (see Section E.4 in the Supplementary Appendix for a detailed explanation). For the choice of the tuning parameter $\alpha$, we choose it for each model $w$ by minimizing the GCV criterion $GCV(\alpha)$.

\subsection{Overall evaluation of GSD}\label{SS:Evaluation}
This subsection presents our empirical results for the three countries/areas and the three economic periods as explained previously to answer questions (A) and (B) in the Section \ref{s:Introduction}. We estimate the following models: (a) the nowcasting models $M_{(1)},\ldots,M_{(13)}$ accounting for the full set of information (GSD, hard data, and soft data) without preselection, (b) the nowcasting models $M_{(1)},\ldots,M_{(13)}$ accounting for the full set of information (GSD, hard data, and soft data) with the preselection of Step 1 (i.e. Ridge after Model selection approach), (c) the nowcasting models $M_{(1)},\ldots,M_{(13)}$ by using only GSD (i.e. $\beta_{s,w} = \beta_{h,w} = 0$ for every $w=1,\ldots,13$ in Equation \eqref{eq:model:weeks}), and (d) the models that only account for hard and soft data (i.e., without GSD, $\beta_{g,w} = 0$ for every $w=1,\ldots,13$ in Equation \eqref{eq:model:weeks}). Root MSFE (RMSFE) results are presented in Tables \ref{Table:Panel:1}--\ref{Table:Panel:9} in Appendix \ref{s:Appendix:paper}, each row corresponding to those four models. The complete tables containing the RMSFEs for six values of $\tau$ are in Section E.7 in the Supplementary Appendix. From there we see how the choice of $\lambda$ impacts the nowcasting accuracy, and for policy-makers, we recommend to use small values of $\tau$ ($1\%$ or $0.5\%$).\\
\indent By looking at rows 2--4 of Tables \ref{Table:Panel:1}--\ref{Table:Panel:9}, we can compare nowcasts obtained with and without GSD in a pseudo real-time exercise to assess: (i) if GSD are informative when no official data are available for the forecaster and (ii) the extent to which GSD remain informative when official data become available.
The first stylized fact that we observe in our results is the downward sloping evolution of RMSFEs throughout the quarter stemming from models $M_{(1)},\ldots,M_{(13)}$ with full information (second row). This result is in line with what could be expected from nowcasting exercises when integrating increasingly more information throughout the quarter (see \cite{Angelinietal11}).
Second, when using only Google information (third row), we still observe a decline of RMSFEs throughout the quarter but to a much lower extent. Third, when focusing on the beginning of the quarter, models that only integrate Google information provide very reasonable RMSFEs, which are slightly higher than those obtained at the end of the quarter. This means that using only Google data at the beginning of the quarter, when no other official information is available about the current state of the economy, is a pertinent strategy for economists aiming at tracking GDP.\\
\indent Let us now focus on the gain from using our \emph{Ridge after Model Selection} strategy. During both calm and sudden shift periods (namely, 2014--2016 and 2017--2018), this estimation strategy applied to the whole data set (hard data, soft data, and GSD) generally tends to provide the lowest RMSFEs (second row of Tables \ref{Table:Panel:1}--\ref{Table:Panel:9} in the Appendix, lowest values in boldface). This result, which is robust over the different countries/areas, does not hold during the recession period (2008--2009). Thus, we have here few important results. First, by comparing the first and second rows of Tables \ref{Table:Panel:1}--\ref{Table:Panel:9}, we can conclude that, outside recession periods, our \emph{Ridge after Model Selection} strategy outperforms a strategy that would skip the data preselection step. Second, when comparing the second, third and fourth rows of Tables \ref{Table:Panel:1}--\ref{Table:Panel:9}, we observe that, outside recession periods, combining information (i.e. macroeconomic and Google data) generally leads to more accurate nowcasts than those based solely on either pure macroeconomic information or pure Google information. These findings are robust to the size of the training sample: we also used a smaller training sample of the same length as the one used for 2008q1--2009q2.\\
\indent Third, recession periods exhibit a very specific pattern because preselecting data does not necessarily generate lower RMSFEs during those phases of the business cycle. Indeed, for almost all weeks within the quarter, the Ridge model that only integrates Google data without preselection outperforms other models (third rows of Tables \ref{Table:Panel:3}, \ref{Table:Panel:6}, \ref{Table:Panel:9}). Thus, our results suggest that during a recession, (i) forecasters do not have to preselect data and (ii) GSD provide more accurate information than official macroeconomic data. This latter result can also be observed during the COVID-19 recession, during which research shows that using new sources of high-frequency data is an effective way to get more reactive and accurate nowcasts of the economic activity (\cite{Lewisetal20}).

\subsection{Robustness checks}\label{ss:robustness:check}
Next, we present the results of two robustness checks we carried out for the EA. The first concerns a true real-time analysis executed by using vintages of GDP and industrial production. The second robustness check consists in controlling for additional official macroeconomic series other than industrial production and opinion surveys.

\subsubsection{A true real-time analysis}\label{ss:real_time_analyses}
We perform the first analysis over the period 2014--2016 for which we use vintages of data for EA GDP and industrial production (survey data are generally not revised) and account for the observed timeline of data release as provided by Eurostat. When available, we also include lagged GDP growth among the explanatory variables of the nowcasting models (see Table 25 in the Supplementary Appendix, which gives the exact weeks in the OOS period 2014q1--2016q1 for which the lagged GDP is included in the real-time analysis).\\
\indent Tables 36--38 in the Supplementary Appendix present the RMSFEs values that compare the same four models (a)--(d) as in the previous pseudo real-time analysis.
Overall, the results that we have illustrated herein for the pseudo real-time exercise still hold in true real-time. In particular, they show that (i) nowcasting accuracy improves throughout the quarter, (ii) GSD provide valuable information at the beginning of the quarter when no official information is available, (iii) combining macroeconomic information with Google information improves the results, and (iv) our Ridge after Model Selection strategy outperforms the other approaches in terms of nowcasting accuracy. These results increase our confidence about the reliability of the real-time use of GSD when nowcasting EA GDP.

\subsubsection{Controlling for additional macroeconomic variables}\label{ss:robustness:check:other:variables}
So far, we have controlled for economic surveys, $S_t$, and industrial production, $IP_t$, as official series. This choice is motivated by the fact that practitioners consider both series the two most important variables to asses the EA economic state. Here, we aim to check the robustness of our evaluation about GSD to a richer macroeconomic information set. For this purpose, we include additional macroeconomic series among the covariates $x_{t,s}^{(w)}$ and $x_{t,h}^{(w)}$ in model \eqref{eq:model:weeks} (see Table 26 in the Supplementary Appendix for a description of the 22 macroeconomic variables used). Studies in the nowcasting literature commonly use these types of macroeconomic series (\textit{e.g.}, sales, exports, unemployment rate) which are continuously monitored by policy makers and market participants.
We refer to this richer set of variables as a {\it Big Official Set} and to the set made of the previously considered variables, $IP_t$ and $S_t$, as a {\it Small Official Set}. The robustness check covers two periods: $2014q1-2016q1$, which exhibited cyclical stability, and $2017q1-2018q4$, which exhibited a downward shift in the EA-GDP growth series.\\
\indent As a performance measure, we take the ratios between the RMSFEs obtained by using GSD together with the {\it Small Official Set} of data (resp. with the {\it Big Official Set} of data) in the numerator (resp. in the denominator). A ratio larger than one indicates that including additional official series improves nowcasting accuracy, and conversely.\\
\indent Results shown in Table 45 of the Supplementary Appendix highlight that in the period of cyclical stability (2014q1--2016q1) the inclusion of additional macroeconomic variables when using our Ridge after Model Selection strategy does not generally improve the nowcasting accuracy except for week 4 and, to a lesser extent week 2. This result still holds when no preselection of data occurs and when we only account for macroeconomic variables. This result provides a strong indication of the robustness of our findings in that it means that using only industrial production and a survey provides a good estimate when nowcasting EA GDP. During a downward shift in the GDP, as in the period $2017q1-2018q4$, it seems worth including a larger set of macroeconomic variables in weeks $4$ and $6$ to $10$ (row $4$ of Table 45). For the first three weeks and last three weeks, the {\it Small Official} data set is preferred. However, if we compare the results obtained without preselection of GSD (row $5$), inclusion of additional macroeconomic variables does not improve the nowcasting accuracy. A possible explanation is that the uncertainty generated by the trade war does not have a strong common impact on all macroeconomic variables and does not adversely affect economic activity across the board but it directly affects the GSD.
% ====================================================================

\section{Conclusions}\label{S:Conclusions}
Large data sets arising from alternative sources of information have gained popularity among macroeconomists whose goal is to assess the current state of the economy on a high-frequency basis. The design of an appropriate econometric methodology for macroeconomic nowcasting must take into account the specific structure of the alternative data used. This paper deals with the particular structure of GSD and proposes an econometric approach for nowcasting macroeconomic quantities based on GSD together with official data when available. Our proposed \emph{Ridge after Model Selection} approach consists of two step: in the first step, GSD variables are preselected, conditionally on the official variables, by targeting the macroeconomic aggregate to be nowcast. In the second step, a Ridge regularization is applied to those preselected GSD and official variables, with Ridge tuning parameter chosen by GCV.\\
\indent Our theoretical contribution consists in showing different types of optimality properties of our proposed procedure. First, we prove that our targeted preselection retains all the active variables in the true model with probability approaching one as $N,T\rightarrow \infty$. Second, we derive an upper bound for both the in-sample and OOS prediction errors associated with the \emph{Ridge after Model Selection} estimator as well as for the parameter estimator. Finally, we are the first to demonstrate optimality of GCV for OOS prediction in the setting of Ridge regularization.\\
\indent We illustrate our procedure through numerical studies and an empirical application in which we nowcast GDP growth rates for the EA as whole, the United States, and Germany by combining standard macroeconomic variables and alternative GSD. Empirical results show that GSD contain valuable information about the current economic state and that combining standard macroeconomic variables with GSD variables is generally fruitful. They also suggest that the preselection step is crucial, as it consistently leads to better outcomes. However, recession periods present specific patterns as, during those phases of the business cycle, models that only contain non-preselected GSD tend to outperform in terms of nowcasting accuracy.

\paragraph{Acknowledgments:} The authors gratefully thank the Editor Jianqing Fan, an Associate Editor, and two anonymous referees for their many constructive comments on the previous version of this paper. The authors also thank Roberto Golinelli, Michele Lenza, Francesca Monti, Giorgio Primiceri, Simon Sheng, Hal Varian, and participants at seminars at Harvard/MIT, European University Institute, EDHEC, and Brunel University, as well as those at the 10th ECB Conference on Macro forecasting with large datasets, Data Day@HEC, and Bocconi-Banque de France alternative datasets conferences, for useful comments. They thank Per Nymand-Andersen (ECB) for sharing the Google data set as well as Dario Buono and Rosa Ruggeri-Cannata (Eurostat) for sending the real-time euro area GDP data. They are grateful to Vivien Chbicheb for outstanding research assistance. An early version of this paper was circulated under the title ``Macroeconomic nowcasting with big data through the lens of a targeted factor model''.  The authors report there are no competing interests to declare.

\paragraph{Funding:}
The second author gratefully acknowledges financial support from Hi!Paris, Europlace Institute of Finance, and ANR-21-CE26-0003, and hospitality from Fondation Banque de France. This research project started while the first author was working for the Banque de France.

\paragraph{Supplementary Material:} The Supplementary material consists of an online Supplementary Appendix that contains additional theoretical results, all the proofs of the results in the paper, additional simulations, figures, tables, and the discussion of macroeconomic series used in the empirical study.
%\clearpage
\begin{spacing}{1}
\setlength{\bibsep}{0.3cm}

\bibliographystyle{alphanat}
\bibliography{AnnaBib}
\end{spacing}

\newpage
\appendix
\section{Appendix}\label{s:Appendix:paper}

\begin{table}[h!]
\resizebox{\columnwidth}{!}{
  \begin{tabular}{|c||c|c|c|c|c|c|}
    \hline
    \hline
    & \multicolumn{6}{|c|}{$\delta = 0.2\iota$}\\
    \hline
    & \multicolumn{2}{|c|}{$N=200$, $T=150$, $s=105$} & \multicolumn{2}{|c||}{$N=200$, $T=150$, $s=110$} & \multicolumn{2}{|c||}{$N=200$, $T=100$, $s=110$}\\
    \hline
    FDR & MSER & MSFER & MSER & MSFER & MSER & MSFER\\
    \hline
    & \multicolumn{6}{|c|}{$\Psi=I_N$}\\
    \hline
    20\% & 0.9526 & 0.8104 & 1.0206 & 0.8478 & 0.7831 & 1.1369\\
    10\% & 0.9620 & 0.8412 & 1.0061 & 0.8863 & 0.9226 & 1.0240\\
    5\% & 0.9414 & 0.8364 & 0.9797 & 0.8794 & 0.9667 & 1.0041 \\
    2.5\% & 0.9343 & 0.8611 & 0.9770 & 0.9030 & 0.9825 & 1.0019\\
    1\% & 0.9130 & 0.8753 & 0.9538 & 0.9220 & 1.0001 & 1.0188\\
    0.5\% & 0.9089 & 0.8781 & 0.9485 & 0.9324 & 1.0013 & 1.0146\\
    \hline
    & \multicolumn{6}{|c|}{$\Psi=\left((0.5)^{|j-k|}\right)_{j,k}$}\\
    \hline
    20\% & 0.9643 & 0.7904 & 1.0066 & 0.8154 & 0.9051 & 1.1521\\
    10\% & 0.9657 & 0.8325 & 1.0053 & 0.8724 & 0.9492 & 1.0860\\
    5\% & 0.9236 & 0.8387 & 0.9654 & 0.8812 & 0.9805 & 1.0222\\
    2.5\% & 0.9178 & 0.8503 & 0.9642 & 0.8905 & 0.9996 & 1.0216\\
    1\% & 0.9011 & 0.8593 & 0.9477 & 0.9025 & 1.0262 & 1.0287\\
    0.5\% & 0.8865 & 0.8607 & 0.9403 & 0.9097 & 1.0301 & 1.0427\\
    \hline
    \hline
  \end{tabular}}
\caption{{\footnotesize Effect of $N,T,s$ on the in-sample and OOS prediction error. In-sample (MSER) and OOS MSE (MSFER) are expressed as ratios with respect to the case $N=150, T=100,s=105$. FDR denotes the percentage of false positives that can be tolerated.}}\label{Table:Results:NTs:paper:1}
\end{table}

\begin{table}[h!]
\resizebox{\columnwidth}{!}{
  \begin{tabular}{|c||c|c|c|c|c|c|}
    \hline
    \hline
    & \multicolumn{6}{|c|}{$\delta = 0.8\iota$}\\
    \hline
    & \multicolumn{2}{|c||}{$N=200$, $T=150$, $s=105$} & \multicolumn{2}{|c|}{$N=200$, $T=150$, $s=110$} & \multicolumn{2}{|c||}{$N=200$, $T=100$, $s=110$}\\
    \hline
    FDR & MSER & MSFER & MSER & MSFER & MSER & MSFER\\
    \hline
    & \multicolumn{6}{|c|}{$\Psi=I_N$}\\
    \hline
    20\% & 0.9481 & 0.8047 & 1.0185 & 0.8433 & 0.7856 & 1.11799\\
    10\% & 0.9476 & 0.8438 & 0.9955 & 0.8885 & 0.9184 & 1.01455\\
    5\% & 0.9275 & 0.8280 & 0.9718 & 0.8747 & 0.9767 & 0.99942\\
    2.5\% & 0.9338 & 0.8477 & 0.9768 & 0.8932 & 1.0001 & 1.00466\\
    1\% & 0.9136 & 0.8666 & 0.9547 & 0.9149 & 0.9993 & 1.02664\\
    0.5\% & 0.9097 & 0.8694 & 0.9487 & 0.9248 & 0.9995 & 1.01358\\
    \hline
    & \multicolumn{6}{|c|}{$\Psi=\left((0.5)^{|j-k|}\right)_{j,k}$}\\
    \hline
    20\% & 0.9488 & 0.7826 & 0.9952 & 0.8103 & 0.8957 & 1.14796\\
    10\% & 0.9511 & 0.8283 & 0.9941 & 0.8686 & 0.9463 & 1.07847\\
    5\% & 0.9081 & 0.8314 & 0.9555 & 0.8779 & 0.9928 & 1.02038\\
    2.5\% & 0.9018 & 0.8368 & 0.9499 & 0.8777 & 1.0081 & 1.01674\\
    1\% & 0.8987 & 0.8490 & 0.9452 & 0.8942 & 1.0303 & 1.03002\\
    0.5\% & 0.8866 & 0.8518 & 0.9373 & 0.9010 & 1.0321 & 1.04643\\
    \hline
    \hline
  \end{tabular}}
\caption{{\footnotesize Effect of $N,T,s$ on the in-sample and OOS prediction error. In-sample (MSER) and OOS MSE (MSFER) are expressed as ratios with respect to the case $N=150, T=100,s=105$. FDR denotes the percentage of false positives that can be tolerated.}}\label{Table:Results:NTs:paper:2}
\end{table}

\begin{singlespace}

\begin{sidewaystable}
\centering
\scalebox{0.9}{
  \begin{tabular}{|l||ccccccccccccc|}
    \hline
    \hline
    \multicolumn{14}{c}{EA -- Nowcasting during $2014q1-2016q1$}\\
    \hline
    & M1 & M2 & M3 & M4 & M5 & M6 & M7 & M8 & M9 & M10 & M11 & M12 & M13 \\
    \hline
    \begin{tabular}{@{}c@{}}
    \footnotesize{Ridge (Google+}\\\footnotesize{S+IP)}\end{tabular} & 0.4467 & 0.4816 & 0.3897 & 0.3659 & 0.3239 & 0.3829 & 0.3901 & 0.3609 & 0.3427 & 0.3422 & 0.3103 & 0.3142 & 0.3111\\
    \begin{tabular}{@{}c@{}}
    \footnotesize{Sel + Ridge (Google+}\\\footnotesize{S+IP)}\end{tabular} & \textbf{0.2889} & \textbf{0.2607} & \textbf{0.2400} & \textbf{0.2493} & \textbf{0.1747} & \textbf{0.1706} & \textbf{0.1695} & \textbf{0.1608} & \textbf{0.1641} & \textbf{0.1668} & 0.2222 & \textbf{0.2178} & 0.2082\\
    \footnotesize{Sel + Ridge (Google)} & 0.3026 & 0.2769 & 0.2841 & 0.3008 & 0.3052 & 0.3107 & 0.3001 & 0.2974 & 0.2984 & 0.2975 & 0.2964 & 0.2867 & 0.2880\\
    \footnotesize{No Google} & &&&& 0.1807 &&&& 0.1897 && \textbf{0.1928} && \textbf{0.2017}\\
    \hline
    \hline
  \end{tabular}}
\caption{\footnotesize{RMSFEs corresponding to the nowcasting period $2014q1-2016q1$. ``Ridge (Google+S+IP)'' refers to model \eqref{eq:model:weeks} with Google data, Survey, and Industrial Production Index (IPI) estimated without pre-selection, ``Sel + Ridge (Google+S+IP)'' refers to model \eqref{eq:model:weeks} with Google data, Survey, and IPI estimated with our \emph{Ridge after Model Selection} procedure, ``Sel + Ridge (Google)'' refers to model \eqref{eq:model:weeks} with only Google data (preselected), and ``No Google'' refers to models without Google data ($NoGoogle_1$ - $NoGoogle_4$ in Table 23 in the Supplementary Appendix).}}\label{Table:Panel:1}

\vspace{2\baselineskip}
\centering
%\begin{adjustbox}{width=\columnwidth,center}
\scalebox{0.9}{
  \begin{tabular}{|l||ccccccccccccc|}
    \hline
    \hline
    \multicolumn{14}{c}{EA -- Nowcasting during $2017q1-2018q4$}\\
    \hline
    & M1 & M2 & M3 & M4 & M5 & M6 & M7 & M8 & M9 & M10 & M11 & M12 & M13 \\
    \hline
    \begin{tabular}{@{}c@{}}
    \footnotesize{Ridge (Google+}\\\footnotesize{S+IP)}\end{tabular} & 0.5592 & 0.5956 & 0.5713 & 0.5642 & 0.3604 & 0.3605 & 0.3186 & 0.3064 & 0.5100 & 0.4836 & 0.4181 & 0.4527 & 0.4788\\
    \begin{tabular}{@{}c@{}}
    \footnotesize{Sel + Ridge (Google+}\\\footnotesize{S+IP)}\end{tabular} & \textbf{0.3505} & \textbf{0.3306} & 0.3341 & \textbf{0.3227} & \textbf{0.2330} & \textbf{0.2664} & \textbf{0.2501} & \textbf{0.2415} & \textbf{0.2720} & \textbf{0.2451} & \textbf{0.1316} & \textbf{0.1340} & \textbf{0.1314}\\
    \footnotesize{Sel + Ridge (Google)} & 0.3760 & 0.3431 & \textbf{0.3262} & 0.3276& 0.3230 & 0.3167 & 0.3051 & 0.2875 & 0.2894 & 0.2856 & 0.2795 & 0.2763 &0.2700\\
    \footnotesize{No Google} & &&&& 0.4340 &&&& 0.4841 && 0.2871 && 0.3177\\
    \hline
    \hline
  \end{tabular}}
\caption{\footnotesize{RMSFEs corresponding to the nowcasting period $2017q1-2018q4$. ``Ridge (Google+S+IP)'' refers to model \eqref{eq:model:weeks} with Google data, Survey, and Industrial Production Index (IPI) estimated without preselection, ``Sel + Ridge (Google+S+IP)'' refers to model \eqref{eq:model:weeks} with Google data, Survey, and IPI estimated with our \emph{Ridge after Model Selection} procedure, ``Sel + Ridge (Google)'' refers to model \eqref{eq:model:weeks} with only Google data (preselected), and ``No Google'' refers to models without Google data ($NoGoogle_1$ - $NoGoogle_4$ in Table 23 in the Supplementary Appendix).}}\label{Table:Panel:2}
%\end{adjustbox}
%
\vspace{2\baselineskip}
\centering
\scalebox{0.9}{
  \begin{tabular}{|l||ccccccccccccc|}
    \hline
    \hline
    \multicolumn{14}{c}{EA -- Nowcasting during recession periods}\\
    \hline
    & M1 & M2 & M3 & M4 & M5 & M6 & M7 & M8 & M9 & M10 & M11 & M12 & M13 \\
    \hline
    \begin{tabular}{@{}c@{}}
    \footnotesize{Ridge (Google+}\\\footnotesize{S+IP)} \end{tabular} & \textbf{1.4601} & \textbf{1.2693} & \textbf{1.2268} & \textbf{1.0596} & 1.0458 & 0.9831 & \textbf{0.9340} & 1.0843 & 1.1047 & 1.1047 & \textbf{0.9632} & \textbf{0.9401} & \textbf{0.9101}\\
    \begin{tabular}{@{}c@{}}
    \footnotesize{Sel + Ridge (Google+}\\\footnotesize{S+IP)} \end{tabular} & 1.5481 & 1.4771 & 1.5257 & 1.6215 & 1.5581 & 1.6184 & 1.6345 & 1.6313 & 1.6344 & 1.6677 & 1.0953 & 1.0468 & 1.0622\\
    \footnotesize{Ridge (Google)} & \textbf{1.4601} & \textbf{1.2693} & \textbf{1.2268} & \textbf{1.0596} & \textbf{0.7745} & \textbf{0.8267} & 1.0072 & \textbf{1.0732} & \textbf{1.0415} & \textbf{1.0042} & 0.9962 & 0.9735 & 0.9657\\
    \footnotesize{No Google} & &&&& 1.5269 &&&& 1.4241 && 1.6351 && 1.2888\\
    \hline
    \hline
  \end{tabular}}
\caption{\footnotesize{RMSFEs corresponding to the nowcasting period $2008q1-2009q2$. ``Ridge (Google+S+IP)'' refers to model \eqref{eq:model:weeks} with Google data, Survey, and Industrial Production Index (IPI) estimated without preselection, ``Sel + Ridge (Google+S+IP)'' refers to model \eqref{eq:model:weeks} with Google data, Survey, and IPI estimated with our \emph{Ridge after Model Selection} procedure, ``Ridge (Google)'' refers to model \eqref{eq:model:weeks} with only Google data estimated without preselection, and ``No Google'' refers to models  without Google data ($NoGoogle_1$ - $NoGoogle_4$ in Table 23 in the Supplementary Appendix).}}\label{Table:Panel:3}
\end{sidewaystable}

%-- US --
\begin{sidewaystable}
    \centering
    \scalebox{0.9}{
      \begin{tabular}{|l||ccccccccccccc|}
        \hline
        \hline
        \multicolumn{14}{c}{U.S. -- Nowcasting during $2014q1-2016q1$}\\
        \hline
        & M1 & M2 & M3 & M4 & M5 & M6 & M7 & M8 & M9 & M10 & M11 & M12 & M13 \\
        \hline
        \begin{tabular}{@{}c@{}}
        \footnotesize{Ridge (Google+}\\\footnotesize{S+IP)}\end{tabular} & 0.5843 & 0.5177 & 0.5771 & 0.5769 & 0.4588 & 0.4735 & 0.4207 & 0.4213 & 0.4315 & 0.4313 & 0.4479 & 0.4488 & 0.4786\\
        \begin{tabular}{@{}c@{}}
        \footnotesize{Sel + Ridge (Google+}\\\footnotesize{S+IP)}\end{tabular} & 0.4889 & \textbf{0.4792} & \textbf{0.4647} & \textbf{0.4670} & 0.4101 & \textbf{0.4277} & \textbf{0.3957} & \textbf{0.3922} & \textbf{0.3948} & \textbf{0.3933} & \textbf{0.4233} & \textbf{0.4273} & 0.4509\\
        \footnotesize{Sel + Ridge (Google)} & \textbf{0.4873} & 0.4833 & 0.4829 & 0.4816 & 0.4777 & 0.4740 & 0.4750 & 0.4751 & 0.4745 & 0.4746 & 0.4753 & 0.4749 & 0.4703\\
        \footnotesize{No Google} &  &  &  &  & \textbf{0.4062} &  & 0.4061 &  & 0.4156 &  & 0.4260 &  & \textbf{0.4466}\\
        \hline
        \hline
      \end{tabular}}
    \caption{\footnotesize{RMSFEs corresponding to the nowcasting period $2014q1-2016q1$. ``Ridge (Google+S+IP)'' refers to model \eqref{eq:model:weeks} with Google data, Survey, and Industrial Production Index (IPI) estimated without preselection, ``Sel + Ridge (Google+S+IP)'' refers to model \eqref{eq:model:weeks} with Google data, Survey, and IPI estimated with our \emph{Ridge after Model Selection} procedure, ``Sel + Ridge (Google)'' refers to model \eqref{eq:model:weeks} with only Google data (preselected), and ``No Google'' refers to models without Google data ($NoGoogle_1$ - $NoGoogle_5$ in Table 24 in the Supplementary Appendix).}}\label{Table:Panel:4}

  \vspace{2\baselineskip}
    \centering
    \scalebox{0.9}{
      \begin{tabular}{|l||ccccccccccccc|}
        \hline
        \hline
        \multicolumn{14}{c}{U.S. -- Nowcasting during $2017q1-2018q4$}\\
        \hline
        & M1 & M2 & M3 & M4 & M5 & M6 & M7 & M8 & M9 & M10 & M11 & M12 & M13 \\
        \hline
        \begin{tabular}{@{}c@{}}
        \footnotesize{Ridge (Google+}\\\footnotesize{S+IP)}\end{tabular} & 0.4746 & 0.4090 & 0.4625 & 0.4549 & 0.2762 & 0.3021 & 0.2401 & 0.2407 & 0.2963 & 0.2506 & 0.2555 & 0.1791 & 0.2081\\
        \begin{tabular}{@{}c@{}}
        \footnotesize{Sel + Ridge (Google+}\\\footnotesize{S+IP)}\end{tabular} & 0.3639 & 0.3601 & \textbf{0.3092} & \textbf{0.3181} & \textbf{0.1735} & \textbf{0.1685} & \textbf{0.1347} & \textbf{0.1330} & \textbf{0.1042} & \textbf{0.0991} & \textbf{0.1187} & \textbf{0.1081} & \textbf{0.1320}\\
        \footnotesize{Sel + Ridge (Google)} & \textbf{0.3482} & \textbf{0.3335} & 0.3177 & 0.3270 & 0.3168 & 0.3103 & 0.3061 & 0.3055 & 0.2949 & 0.2833 & 0.2770 & 0.2816 & 0.2757\\
        \footnotesize{No Google} & &  &  &  & 0.2598 &  & 0.2255 &  & 0.3604 &  & 0.3510 &  & 0.2979\\
        \hline
        \hline
      \end{tabular}}
    \caption{\footnotesize{RMSFEs corresponding to the nowcasting period $2017q1-2018q4$. ``Ridge (Google+S+IP)'' refers to model \eqref{eq:model:weeks} with Google data, Survey, and Industrial Production Index (IPI) estimated without preselection, ``Sel + Ridge (Google+S+IP)'' refers to model \eqref{eq:model:weeks} with Google data, Survey, and IPI estimated with our \emph{Ridge after Model Selection} procedure, ``Sel + Ridge (Google)'' refers to model \eqref{eq:model:weeks} with only Google data (preselected), and ``No Google'' refers to models without Google data ($NoGoogle_1$ - $NoGoogle_5$ in Table 24 in the Supplementary Appendix).}}\label{Table:Panel:5}
  \vspace{2\baselineskip}
    \centering
    \scalebox{0.9}{
      \begin{tabular}{|l||ccccccccccccc|}
        \hline
        \hline
        \multicolumn{14}{c}{U.S. -- Nowcasting during recession periods}\\
        \hline
        & M1 & M2 & M3 & M4 & M5 & M6 & M7 & M8 & M9 & M10 & M11 & M12 & M13 \\
        \hline
        \begin{tabular}{@{}c@{}}
        \footnotesize{Ridge (Google+}\\\footnotesize{S+IP)} \end{tabular} & \textbf{1.0156} & 1.0507 & \textbf{1.0506} & \textbf{1.0282} & 1.0967 & 1.1191 & 1.1185 & 1.1986 & 1.2123 & 1.1884 & 1.1738 & 1.1801 & 1.1217\\
        \begin{tabular}{@{}c@{}}
        \footnotesize{Sel + Ridge (Google+}\\\footnotesize{S+IP)} \end{tabular} & 1.0320 & 1.0611 & 1.0590 & 1.0521 & 1.2030 & 1.2029 & 1.1929 & 1.1893 & 1.1815 & 1.1751 & 1.1319 & 1.1307 & 1.0396\\
        \footnotesize{Ridge (Google)} & \textbf{1.0156} & 1.0507 & \textbf{1.0506} & \textbf{1.0282} & 0.9744 & \textbf{0.9320} & \textbf{0.9731} & \textbf{0.9991} & \textbf{1.0061} & \textbf{1.0158} & \textbf{1.0204} & \textbf{1.0196} & 1.2224\\
        \footnotesize{No Google} & &  &  &  & \textbf{0.8439} &  & 1.1286 &  & 1.0580 &  & 1.0659 &  & \textbf{0.7828}\\
        \hline
        \hline
      \end{tabular}}
    \caption{\footnotesize{RMSFE corresponding to the nowcasting period $2008q1-2009q2$. ``Ridge (Google+S+IP)'' refers to model \eqref{eq:model:weeks} with Google data, Survey, and Industrial Production Index (IPI) estimated without preselection, ``Sel + Ridge (Google+S+IP)'' refers to model \eqref{eq:model:weeks} with Google data, Survey, and IPI estimated with SIS pre-selection, ``Ridge (Google)'' refers to model \eqref{eq:model:weeks} with only Google data estimated without pre-selection, and ``No Google'' refers to models without Google data ($NoGoogle_1$ - $NoGoogle_5$ in Table 24 in the Supplementary Appendix).}}\label{Table:Panel:6}
\end{sidewaystable}

%-- Germany --
\begin{sidewaystable}
    \centering
    \scalebox{0.9}{
      \begin{tabular}{|l||ccccccccccccc|}
        \hline
        \hline
        \multicolumn{14}{c}{Germany -- Nowcasting during $2014q1-2016q1$}\\
        \hline
        & M1 & M2 & M3 & M4 & M5 & M6 & M7 & M8 & M9 & M10 & M11 & M12 & M13 \\
        \hline
        \begin{tabular}{@{}c@{}}
        \footnotesize{Ridge (Google+}\\\footnotesize{S+IP)}\end{tabular} & 0.3316 & 0.3225 & 0.3296 & 0.3365 & 0.3137 & 0.2905 & 0.2687 & 0.2690 & 0.2695 & 0.2684 & 0.2713 & 0.2754 & 0.2698\\
        \begin{tabular}{@{}c@{}}
        \footnotesize{Sel + Ridge (Google+}\\\footnotesize{S+IP)}\end{tabular} & 0.2619 & 0.2454 & \textbf{0.2219} & \textbf{0.2306} & \textbf{0.2378} & 0.2406 & 0.2373 & 0.2382 & 0.2429 & 0.2460 & 0.2717 & 0.2794 & 0.2754\\
        \footnotesize{Sel + Ridge (Google)} & \textbf{0.2265} & \textbf{0.2266} & 0.2484 & 0.2436 & 0.2433 & \textbf{0.2341} & \textbf{0.2310} & \textbf{0.2296} & \textbf{0.2362} & \textbf{0.2372} & \textbf{0.2380} & \textbf{0.2470} & \textbf{0.2453}\\
        \footnotesize{No Google} &  &  &  &  & 0.3977 &  &  &  & 0.4208 &  & 0.2914 &  & 0.4325\\
        \hline
        \hline
      \end{tabular}}
    \caption{\footnotesize{RMSFEs corresponding to the nowcasting period $2014q1-2016q1$. ``Ridge (Google+S+IP)'' refers to model \eqref{eq:model:weeks} with Google data, Survey, and Industrial Production Index (IPI) estimated without preselection, ``Sel + Ridge (Google+S+IP)'' refers to model \eqref{eq:model:weeks} with Google data, Survey, and IPI estimated with  our \emph{Ridge after Model Selection} procedure, ``Sel + Ridge (Google)'' refers to model \eqref{eq:model:weeks} with only Google data (preselected), and ``No Google'' refers to models without Google data ($NoGoogle_1$ - $NoGoogle_4$ in Table 23 in the Supplementary Appendix).}}\label{Table:Panel:7}

  \vspace{2\baselineskip}
    \centering
    \scalebox{0.9}{
      \begin{tabular}{|l||ccccccccccccc|}
        \hline
        \hline
        \multicolumn{14}{c}{Germany -- Nowcasting during $2017q1-2018q4$}\\
        \hline
        & M1 & M2 & M3 & M4 & M5 & M6 & M7 & M8 & M9 & M10 & M11 & M12 & M13 \\
        \hline
        \begin{tabular}{@{}c@{}}
        \footnotesize{Ridge (Google+}\\\footnotesize{S+IP)}\end{tabular} & 0.4889 & 0.4894 & 0.4837 & 0.4777 & 0.4352 & 0.4800 & 0.4649 & 0.4441 & 0.5038 & 0.4927 & 0.4382 & 0.4059 & 0.4181\\
        \begin{tabular}{@{}c@{}}
        \footnotesize{Sel + Ridge (Google+}\\\footnotesize{S+IP)}\end{tabular} & 0.3814 & \textbf{0.3794} & \textbf{0.3751} & 0.3769 & 0.3831 & 0.3826 & 0.3876 & 0.3898 & 0.3238 & 0.3181 & 0.3141 & \textbf{0.3088} & \textbf{0.2917}\\
        \footnotesize{Sel + Ridge (Google)} & \textbf{0.3812} & 0.3800 & 0.3788 & \textbf{0.3757} & \textbf{0.3711} & \textbf{0.3680} & \textbf{0.3712} & \textbf{0.3584} & \textbf{0.3184} & \textbf{0.3097} & \textbf{0.3141} & 0.3187 & 0.3140\\
        \footnotesize{No Google} & &  &  &  & 0.6532 &  &  &  & 0.6241 &  & 0.3632 &  & 0.6433\\
        \hline
        \hline
      \end{tabular}}
    \caption{\footnotesize{RMSFEs corresponding to the nowcasting period $2017q1-2018q4$. ``Ridge (Google+S+IP)'' refers to model \eqref{eq:model:weeks} with Google data, Survey, and Industrial Production Index (IPI) estimated without preselection, ``Sel + Ridge (Google+S+IP)'' refers to model \eqref{eq:model:weeks} with Google data, Survey, and IPI estimated with our \emph{Ridge after Model Selection} procedure, ``Sel + Ridge (Google)'' refers to model \eqref{eq:model:weeks} with only Google data (preselected), and ``No Google'' refers to models without Google data ($NoGoogle_1$ - $NoGoogle_4$ in Table 23 in the Supplementary Appendix).}}\label{Table:Panel:8}
  \vspace{2\baselineskip}
    \centering
    \scalebox{0.9}{
      \begin{tabular}{|l||ccccccccccccc|}
        \hline
        \hline
        \multicolumn{14}{c}{Germany -- Nowcasting during recession periods}\\
        \hline
        & M1 & M2 & M3 & M4 & M5 & M6 & M7 & M8 & M9 & M10 & M11 & M12 & M13 \\
        \hline
        \begin{tabular}{@{}c@{}}
        \footnotesize{Ridge (Google+}\\\footnotesize{S+IP)} \end{tabular} & \textbf{1.9279} & \textbf{1.9100} & \textbf{1.9123} & \textbf{1.9212} & 2.2660 & 2.3237 & 2.3365 & 2.3162 & 2.2939 & 2.2790 & 2.7585 & 2.7372 & 2.6581\\
        \begin{tabular}{@{}c@{}}
        \footnotesize{Sel + Ridge (Google+}\\\footnotesize{S+IP)} \end{tabular} & 1.9289 & 1.8990 & 1.9053 & 1.9018 & 2.1467 & 2.1477 & 2.1469 & 2.1461 & 2.1353 & 2.1382 & 2.6461 & 2.6538 & 2.5886\\
        \footnotesize{Ridge (Google)} & \textbf{1.9279} & \textbf{1.9100} & \textbf{1.9123} & \textbf{1.9212} & \textbf{1.9477} & \textbf{2.0123} & \textbf{2.0200} & \textbf{1.9972} & \textbf{1.9793} & \textbf{1.9603} & \textbf{1.9440} & \textbf{1.9173} & \textbf{1.9054}\\
        \footnotesize{No Google} &  &  &  &  & 2.1580 &  &  &  & 2.0660 &  & 2.6107 &  & 1.9803\\
        \hline
        \hline
      \end{tabular}}
    \caption{\footnotesize{RMSFEs corresponding to the nowcasting period $2008q1-2009q2$. ``Ridge (Google+S+IP)'' refers to model \eqref{eq:model:weeks} with Google data, Survey, and Industrial Production Index (IPI) estimated without preselection, ``Sel + Ridge (Google+S+IP)'' refers to model \eqref{eq:model:weeks} with Google data, Survey, and IPI estimated with our \emph{Ridge after Model Selection} procedure, ``Ridge (Google)'' refers to model \eqref{eq:model:weeks} with only Google data estimated without preselection, and ``No Google'' refers to models without Google data ($NoGoogle_1$ - $NoGoogle_4$ in Table 23 in the Supplementary Appendix).}}\label{Table:Panel:9}
\end{sidewaystable}

\end{singlespace}

\end{document}